\documentclass{emulateapj}

\usepackage{amsmath}
\usepackage{array}
\usepackage{bm}
\usepackage{cases}
\usepackage{txfonts}
\usepackage{tipa}
\usepackage{amsfonts}
\usepackage{amssymb}
\usepackage{graphicx}
\usepackage{multirow}
\usepackage{float}
\usepackage{longtable}

\begin{document}

\title{Existence of the Metal-Rich Stellar Halo and High-velocity Thick Disk in the Galaxy }

\author{Yepeng Yan\altaffilmark{1,3}, Cuihua Du\altaffilmark{2}, Hefan Li\ \altaffilmark{1}, Jianrong Shi\altaffilmark{4,2}, Jun Ma\altaffilmark{4,2},Heidi Jo Newberg\altaffilmark{5}}

\affil{$^{1}$School of Physical Sciences, University of Chinese Academy of Sciences, Beijing 100049, China\\
        $^{2}$College of Astronomy and Space Sciences, University of Chinese Academy of Sciences, Beijing 100049, China; ducuihua@ucas.ac.cn \\ 
	$^{3}$Department of Astronomy, Beijing Normal University, Beijing 100875, China\\         
	$^{4}$Key Laboratory of Optical Astronomy, National Astronomical Observatories, Chinese Academy of Sciences, Beijing 100012, China\\ 
        $^{5}$Department of Physics, Applied Physics and Astronomy, Rensselaer Polytechnic Institute, Troy, NY 12180, USA\\   
         }

\begin{abstract}
 Based on the second Gaia data release (DR2), combined with the LAMOST and APOGEE spectroscopic surveys, 
we study the kinematics and metallicity distribution of the high-velocity stars that have a relative speed of at least 220 ${\rm km\ s^{-1}}$ with respect to the local standard of rest in the Galaxy.  The rotational velocity distribution of the high-velocity stars with [Fe/H] $>-1.0$ dex can be well described by a two-Gaussian model, with peaks at $V_{\phi}\sim +164.2\pm0.7$ and  $V_{\phi}\sim +3.0\pm0.3$ ${\rm km\ s^{-1}}$, associated with the thick disk and halo, respectively. This implies that there should exist a high-velocity thick disk (HVTD) and a metal-rich stellar halo (MRSH) in the Galaxy.   The HVTD stars have the same position as the halo in the Toomre diagram and but show the same  rotational velocity and metallicity as the canonical thick disk.  The MRSH stars have  basically the same  rotational velocity,  orbital eccentricity, and position in the Lindblad and Toomre diagram as the canonical halo stars, but they are more metal-rich.  
Furthermore, the metallicity distribution function (MDF) of our sample stars
are well fitted by a four-Gaussian model,  associated with the outer-halo, inner-halo, MRSH, and HVTD, respectively. Chemical and kinematic properties and age imply that the MRSH and HVTD stars may form in situ.   
\end{abstract}

\section{Introduction}
\label{sec:introduction}
The Galactic stellar halo is an important component for unraveling our Galaxy's formation and evolution history. It has been characterized by an old population, metal-poor, high-velocity, large random motions, little if any rotation, and a spheroidal to spherical spatial distribution \citep{Bland-Hawthorn16}. In recent decades, the evidence for the dual halo, namely inner-halo and outer halo, has been found in some studies \citep{Carollo07,Carollo10,deJong10,Beers12,Kinman12,Chen14}. 
The two components differ in their spatial distribution and metallicity. In the spatial distribution, the inner halo component dominates at distances up to $10-15$ kpc from the Galactic center, while the outer-halo component dominates in the region beyond $15-20$ kpc \citep{Carollo07}. The mean metallicity of the inner-halo range from  [Fe/H] $\sim$ $-1.2$ to $-1.7$, while the outer-halo is from [Fe/H] $\sim$ $-1.9$ to $-2.3$ \citep[e.g.,][]{Carollo07,An13,An15,Zuo17,Gu15,Gu16,Gu19,Liu18}, which may depend on the distance of the sample stars.
\par 
The halo population of the Milky Way preserves the fossil record of the formation and evolution of our Galaxy. That record can be accessed through the collection of precision chemical and kinematic information for large samples of halo stars. Two chemically distinct stellar populations: an older, high-$\alpha$, and a younger, low-$\alpha$ halo population,  were also detected \citep[e.g.,][]{Nissen10,Nissen11,Bergemann17,Hayes18,Fernandez-Alvar18}. The kinematic and chemical properties of those stellar populations suggested possible dual formation scenarios for the Galactic halo, comprising in situ star formation as well as accretion from satellite galaxies. 
\par 
Recently, many works have provided multiple  evidences for the existence of a massive ancient merger that provides the bulk of the stars in the inner halo. For example, a broken radius of 20-30 kpc in the Galactic halo, beyond which the stellar density drops precipitously, has been found \citep{Watkins09,Deason11,Sesar11}. \cite{Deason13} argued that the existence of the break radius could be interpreted as the apocenter pileup of the tidal debris from a small number of significant mergers. Additionally, \cite{Belokurov18} showed the velocity ellipsoid  becomes strongly anisotropic for  the halo stars with $-1.7 <$[Fe/H]$<-1.0$, and local velocity distribution appears highly stretched in the radial direction, taking a sausage-like shape, and they suggested that such orbital configurations could show that most of the inner halo stars should be dominated by stars accreted from an ancient (8-11 Gyr ago) massive ($\sim 10^{11} M_{\odot}$)  merger event. This merger event is referred to as the Gaia-Sausage merger \citep[e.g.,][]{Myeong18,Deason18,Lancaster19}. 
\cite{Helmi18} also demonstrated that the inner halo is dominated by debris from the merger of a dwarf galaxy with a mass similar to that of the Small Magellanic Cloud 10 Gyr ago, and the dwarf galaxy referred as Gaia-Enceladus. \cite{Mackereth19b} used APOGEE-DR14 and Gaia-DR2 data sets to show that most nearby halo stars have high orbital eccentricities ($e \gtrsim 0.8$), and chemical trends are similar to current massive dwarf galaxy satellites. These suggested that this population is likely the progeny of a single, massive accretion event that occurred early in the history of the Galaxy, which is consistent with the results of \cite{Belokurov18} and \cite{Helmi18}. By comparing chemo-dynamics of high-$e$ stars with the EAGLE suite of cosmological simulations, \cite{Mackereth19b} constrained such an accreted satellite mass to  $10^{8.5} \lesssim M_{\star} \lesssim 10^{9} M_{\odot}$ and the accretion event likely happened at redshift $z \lesssim 1.5$. Recently, \cite{Myeong19} reported a second substantial accretion episode, it referred as Sequoia Event, and distinct from the Gaia-Sausage (sometimes also referred as Gaia-Enceladus). The Sequoia Event provided the bulk of the high energy retrograde stars in the stellar halo, as well as the recently discovered globular cluster FSR 1758 \citep{Barba19}. The Sequoia stars have lower metallicity by $\sim 0.3$ dex than the Sausage. The Sausage and the Sequoia galaxies could have been associated and accreted at a comparable epoch. All these observation or simulation results led to a claim that accreted stars have been suggested to be the dominant inner halo component.

\par 
Although most previous studies showed that the majority of halo stars have [Fe/H] $<-1$ dex, some recent studies also reported that a large number of metal-rich stars ([Fe/H] $>-1$ dex)  strongly differ from disk stars in kinematics, and instead exhibit halo-like motions \citep[e.g.,][]{Nissen10,Nissen11,Schuster12,Bonaca17,Posti18,Fernandez-Alvar19}. According to kinematic properties of these  metal-rich stars, they are identified as metal-rich halo stars. 
Furthermore, several works have been made to reveal the origin of the metal-rich halo stars. An in-situ metal-rich halo has been corroborated by some works. For example, \cite{Hawkins15} used a sample of 57 high-velocity stars from the fourth data release of the Radial Velocity Experiment to report the discovery of a metal-rich halo star that has likely been dynamically ejected into the halo from the thick disc, which support the theory of \cite{Purcell10} that massive accretion events are believed to heat more metal-rich disk stars so that they are ejected into the halo. 
\cite{Bonaca17} reported that metal-rich halo stars in the solar neighborhood actually formed in situ, rather than having been accreted from satellite systems, based on kinematically identifying halo stars within 3 kpc from the Sun. 

\par
Some studies also \citep{Haywood18,Di Matteo18,Gallart19} detected a substantial population of  with thick disc chemistry on halo-like orbits, and corroborated  their  in-situ origin. \cite{Belokurov19} used a  Gaia DR2  and auxiliary spectroscopy data sets to identify a large population of metal-rich ([Fe/H]$> -0.7$) stars on high eccentric orbits in the rotational velocity versus metallicity plane, and dub the Splash stars.  They confirmed that the Splash stars are predominantly old, but not as old as the stars deposited into the Milky Way in the last major merger, suggesting that the Splash stars could have been born in the Milky Way's  proto-disc prior to the massive ancient accretion event which drastically altered their orbits.  Although these metal-rich halo stars have been found, whether they are a part of halo stars still needs more detailed research.
\par
To understand the complex structure of the Galaxy, we need more information such as chemical abundance and kinematics of large number of individual stars.  The ongoing Large Sky Area Multi-Object Fiber Spectroscopic Telescope survey \citep[LAMOST, also called Guoshoujing Telescope;][]{Zhao12} has released more than five million stellar spectra with stellar parameters in the DR5 catalog. Furthermore, more accurate elemental abundances and radial velocity from high-resolution spectra are provided by
the Apache Point Observatory Galactic Evolution Experiment  \citep[APOGEE;][]{Majewski17} survey. The accurate  kinematic  information requires accurate proper motions and parallaxes with sufficiently small uncertainties, which are provided by the second Gaia data release of Gaia survey \citep{Gaia18a, Gaia18b}. These data sets allow us to explore  the Galactic structure accurately.

\par
In this work, we use a low-resolution sample from the LAMOST DR5 and a high-resolution sample from the APOGEE DR14  combined with the Gaia DR2 to study the metal-rich halo stars kinematically, prove the existence of high-velocity thick disk, and measure  their MDFs. 
The paper is structured as follows: Sect.\ref{sec:data} introduces the observation data, determines the distance and velocity of sample stars, and describes the sample selection. Sect.\ref{sec:kinematics} presents the kinematic evidence for the existence of metal-rich halo stars and high-velocity thick disk, and studies their  kinematic properties. In Sect.\ref{sec:metallicity}, we present metallicity distribution functions (MDFs) of the metal-rich stellar halo and high-velocity thick disk. Sect.\ref{sec:discussion} discusses their potential origins. The summary and conclusions are given in Sect.\ref{sec:summary}.

\section{Data}
\label{sec:data}
\subsection{LAMOST, APOGEE, and Gaia}
\par
Large Sky Area Multi-Object Fiber Spectroscopic Telescope (LAMOST) is a reflecting Schmidt telescope located at Xinglong station, which is operated by National Astronomical Observatories, Chinese Academy of Sciences (NAOC). LAMOST has an effective aperture of 3.6 - 4.9 m in diameter, a focal length of 20 m and 4000 fibers within a field of view of $5^\circ$, which enable it to take 4000 spectra in a single exposure to a limiting magnitude as faint as $r = 19$ (where $r$ denotes magnitude in the SDSS $r$-band) at resolution R = 1800. Its observable sky covers $-10^\circ \sim +90^\circ$ declination and observed wavelength range spans 3,700 {\AA} $\sim$ 9,000  {\AA} \citep{Cui12, Zhao12}.  In this work, we use the LAMOST DR5 catalog that contains over 5 million A-, F-, G-, and K-type stars. Stellar parameters, including radial velocity, effective temperature, surface gravity, and metallicity ([Fe/H]), are delivered from the spectra with the LAMOST Stellar Parameter Pipeline \citep[LASP;][]{Wu11,Luo15}. The accuracy of LASP was tested by selecting 771 stars from the LAMOST commissioning database, and comparing it with the SDSS/SEGUE Stellar Parameter Pipeline (SSPP). The precisions of effective temperature, surface gravity, and metallicity ([Fe/H]) were found to be 167 K, 0.34 dex, and 0.16 dex, respectively.
\begin{figure}[]
	\includegraphics[width=1.0\hsize]{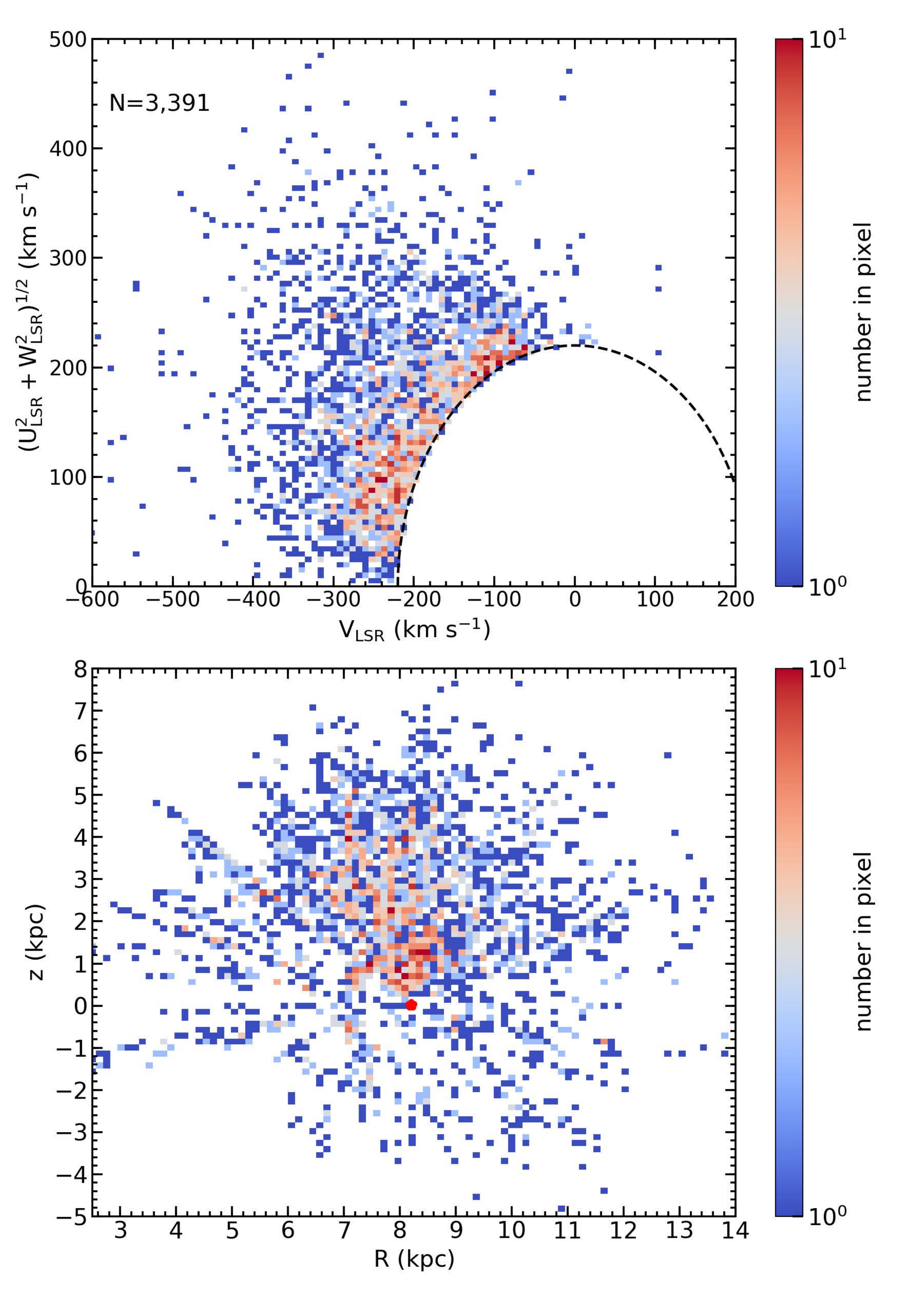}
	\caption{Top panel: Toomre diagram of our high-velocity sample stars  for the high-resolution sample. The black dashed line represents the total spatial velocity  $v_{{\rm tot}} =220$ ${\rm km\ s^{-1}}$, and we adopt $v_{\rm LSR}=232.8\ {\rm km\ s^{-1}}$. Our high-velocity sample stars are defined as $v_{{\rm tot}}>220$ ${\rm km\ s^{-1}}$. Bottom panel: the spatial distribution in cylindrical Galactic coordinates of these high-velocity sample  stars. Red dots indicate the Sun, which is located at ($x_{\odot}$, $y_{\odot}$, $z_{\odot}$) $=$ ($-8.2, 0, 0.015$) kpc. $N$ represents the number of stars.}
	\label{figure1}
\end{figure}

\begin{figure}[]
	\includegraphics[width=1.0\hsize]{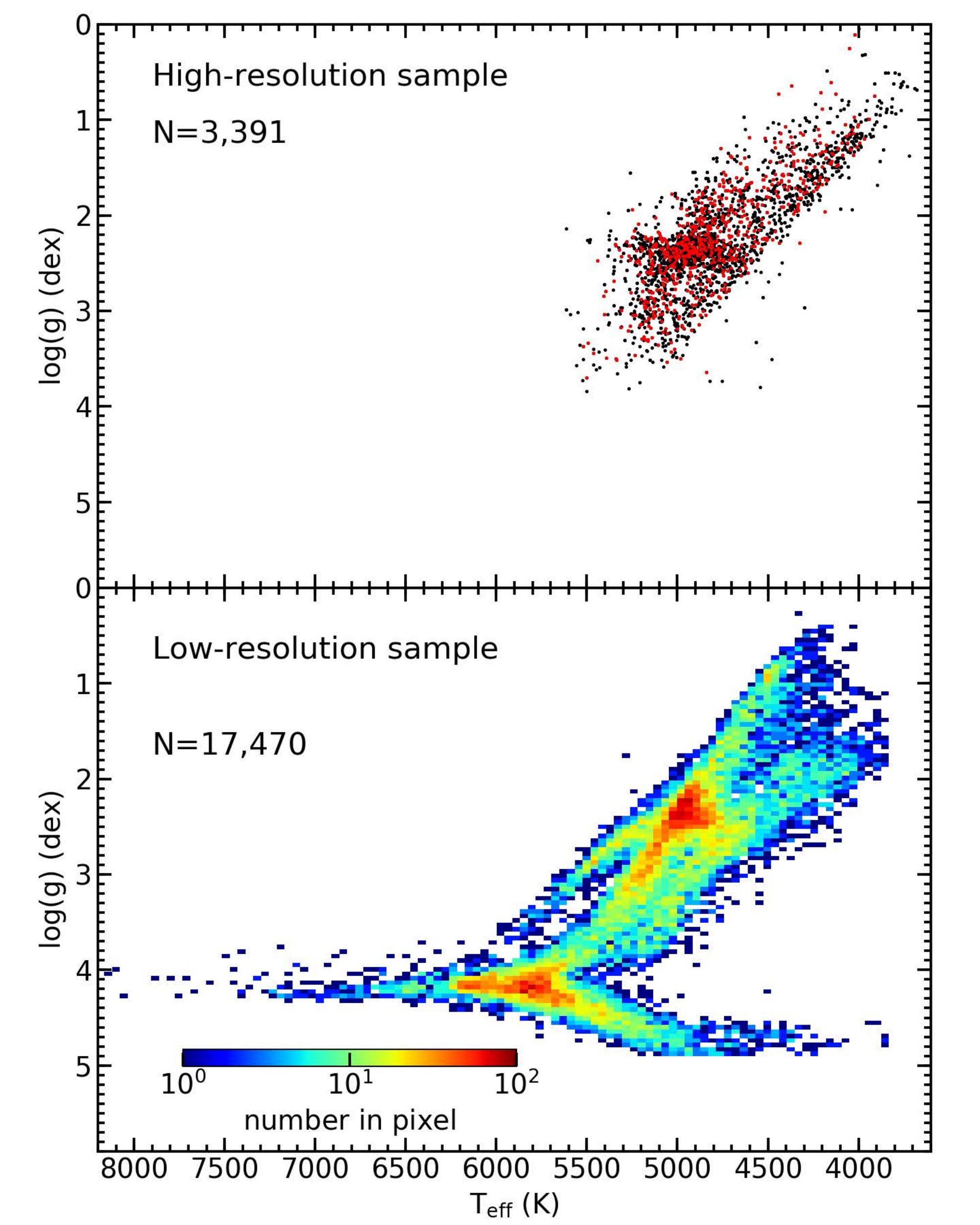}
	\caption{Effective temperature ($T_{\rm eff}$) versus surface gravity (log(g)) diagram for our high-resolution sample (top panel) and low-resolution sample (bottom panel). The sample stars have been selected using  $v_{{\rm tot}} > 220$ ${\rm km\ s^{-1}}$ and given sample criteria. The red dots in the top panel represent 695 common  stars between high-resolution and low-resolution samples. The color bar in the bottom panel represents the number of stars. }
	\label{figure2}
\end{figure}
\par 
The Apache Point Observatory Galactic Evolution Experiment (APOGEE), part of the Sloan Digital Sky Survey III, is a near-infrared (H-band; 1.51-1.70 ${\rm \mu m}$) and high-resolution (R $\sim$ 22,500) spectroscopic survey targeting primarily red giant (RG) stars \citep{Zasowski13}. It provides accurate ($\sim$ 0.1 ${\rm km\ s^{-1}}$) radial velocity, stellar atmospheric parameters, and precise ($\lesssim 0.1$ dex) chemical abundances for about 15 chemical species \citep{Nidever15}. Detailed information about the APOGEE Stellar Parameter and Chemical Abundances Pipeline (ASPCAP) can be found in \cite{Holtzman15} and \cite{Garcia16}.
\par 
Gaia is an ambitious mission to chart a three-dimensional map of the Milky Way, and launched by the European Space Agency (ESA) in 2013. The second Gaia data release, Gaia DR2, provides high-precision positions, parallaxes, and proper motions for 1.3 billion sources brighter than magnitude $G \sim 21$ mag as well as line-of-sight velocities for 7.2 million stars brighter than $G_{RVS}$ = 12 mag \citep{Gaia18a, Gaia18b}. More detailed information about Gaia can be found in \cite{Gaia Collaboration16,Gaia18a, Gaia18b}. 

\begin{figure}[]
	\includegraphics[width=1.0\hsize]{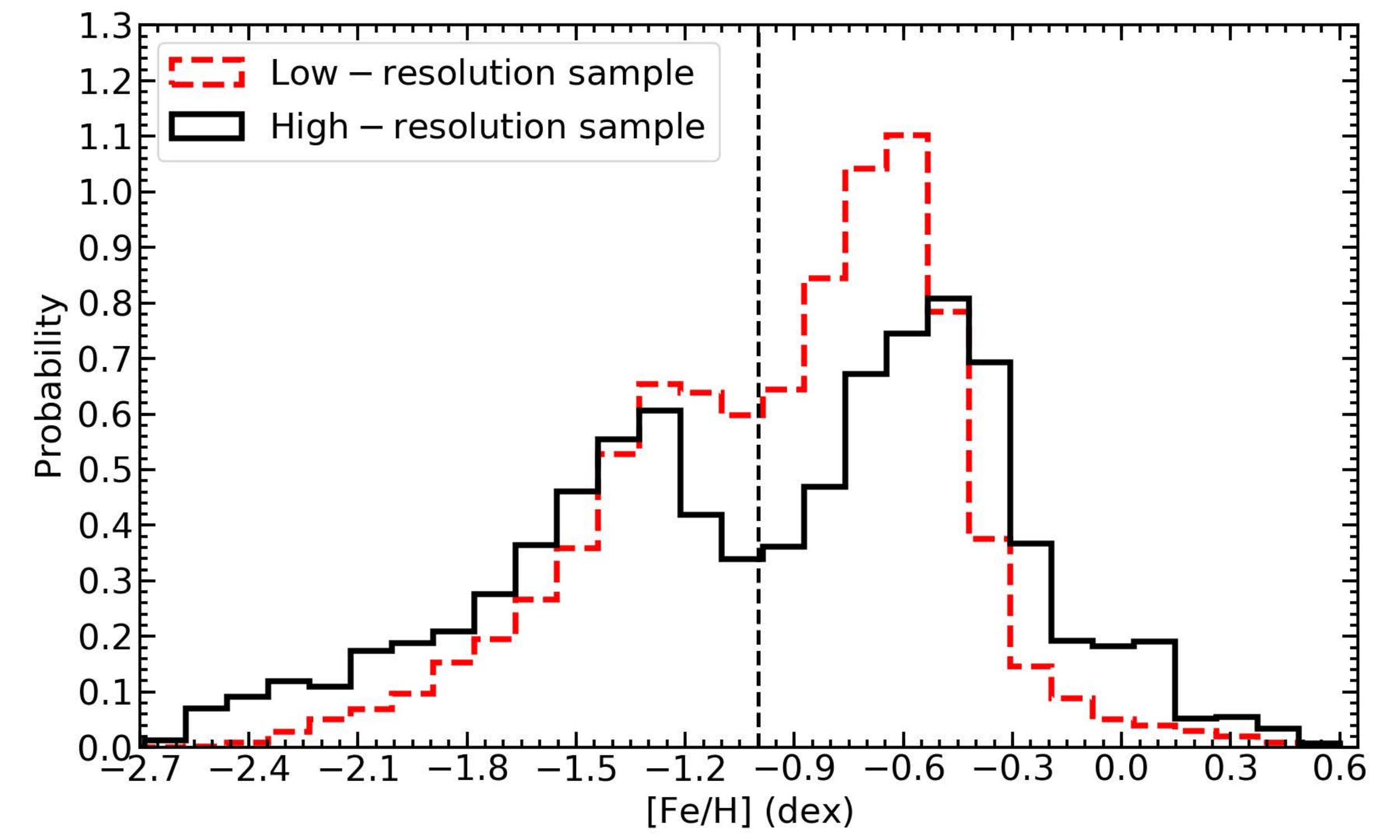}
	\caption{The metallicity distribution of our high-velocity sample stars ($v_{{\rm tot}}>220$ ${\rm km\ s^{-1}}$). The red dashed line represents the low-resolution sample and black line for the high-resolution sample.}
	\label{figure3}
\end{figure}
\subsection{Distance and velocity determination}
\label{sec:Distance}
\par 
In this study, we use two initial samples. One is the low-resolution sample obtained by cross-matching between the LAMOST DR5  and Gaia DR2 catalog, and it can provide a large quantity stars
to study the Galactic disk and halo statistically. In this sample, the stellar parameters such as  [Fe/H], radial velocity, effective temperature, and surface gravity are from the LAMOST DR5 catalog, and proper motion and parallax are from the Gaia DR2 catalog. Another is the high-resolution sample from the APOGEE DR14 and Gaia DR2 catalog, and its stellar parameters ([Fe/H], radial velocity, effective temperature, and surface gravity) are from the APOGEE DR14 catalog,  proper motion, and parallax from the Gaia DR2 catalog. We restrict relative parallax uncertainties  smaller than 20$\%$, the error of proper motion smaller than $0.2$ mas/year, radial velocity uncertainties smaller than $10$ ${\rm km\ s^{-1}}$, the error of [Fe/H] smaller than $0.2$ dex, and signal-to-noise S/N $> 20$ in the $g$-band. \textbf{We also restrict the error of the effective temperature smaller than 150 K and the error of the surface gravity  smaller than 0.3 dex for the low-resolution sample.}

\par 
\cite{Bailer-Jones15} discussed that the inversion of the parallax to obtain distance is not appropriate when the relative parallax error is above 20 percent. Therefore, we  discuss separately the derivation of distances and velocities with $\sigma_{\varpi}/(\varpi-\varpi_{\rm zp}) <0.1$ and $\sigma_{\varpi}/(\varpi-\varpi_{\rm zp}) \geq 0.1$ \citep{Marchetti19}.
The quantity $\varpi$ and $\sigma_{\varpi}$ denote stellar parallax and its error and $\varpi_{\rm zp}$ is the global parallax zero-point of the Gaia observations. \cite{Butkevich17} confirmed that due to various instrumental effects of the Gaia satellite, in particular, to a certain kind of basic-angle variations, these can bias the parallax zero point of an astrometric solution derived from observations. This global parallax zero-point was determined in \cite{Lindegren18} based on observations of quasars: $\varpi_{\rm zp}=-0.029$ mas. Thus, it is necessary to subtract parallax zero-point ($\varpi_{\rm zp}$) when parallax is used to calculate astrophysical quantities \citep{Li19}. 	
For the sample stars with  $\sigma_{\varpi}/(\varpi-\varpi_{\rm zp}) <0.1$, we use simple inversion to calculate  the distance, but for the $\sigma_{\varpi}/(\varpi-\varpi_{\rm zp}) \geq 0.1$ stars, we adopt the Bayesian approach to derive it.  Using the Bayesian approach  to estimate the distance and velocity of the  sample stars will be introduced in the Appendix, as well as the comparisons of our distances and velocities with other works.
Here we only introduce the  distance and velocity determination of  the sample stars with $\sigma_{\varpi}/(\varpi-\varpi_{\rm zp}) <0.1$ using parallax,  proper motion in right ascension ($\mu_{\alpha^*}$) and declination ($\mu_{\delta}$), and radial velocity ($rv$).

\par 
We calculate the Galactocentric Cartesian ($x, y, z$) coordinates from the Galactic ($l, b$) coordinates, and $l$ and $b$ are the Galactic longitude and latitude. We apply a right-handed Galactic-centered Cartesian coordinate with the $x$-axis pointing toward the Galactic center:
\begin{align}
& x = d\,\cos(l)\,\cos(b)- x_\odot  \nonumber \\
& y =  d\,\sin(l)\,\cos(b) \\
& z =  d\,\sin(b). \nonumber
\end{align}
Here, we adopt the distance from  the Sun to the Galactic center is $x_{\odot} = -8.2$ kpc and height above the plane $z_{\odot} = 15$ pc \citep{Bland-Hawthorn16}. In such a coordinate system, the Sun is located at ($x_{\odot}$, $y_{\odot}$, $z_{\odot}$) $=$ ($-8.2, 0, 0.015$) kpc, and $d$ is the distance from the Sun.  The proper motions together with the radial velocity are used to derive the Galactic velocity components $(U, V, W)$ using a right-handed Cartesian coordinate. The directions of $U$ and $W$ are toward the Galactic center and the north Galactic pole, and $V$ is in the direction of the Galactic rotation. The Galactic velocity is relative to  Local Standard of Rest (LSR): $(U, V, W)$ $=$ $(U_{\rm LSR}, V_{\rm LSR}, W_{\rm LSR})$. 
The velocity components in the Galactocentric Cartesian Coordinates can be obtained: $(V_x, V_y, V_z)$ $=$ $(U, V+v_{\rm LSR}, W)$ and $v_{\rm LSR}$ is the LSR velocity, we adopt $v_{\rm LSR}=232.8\ {\rm km\ s^{-1}}$ \citep{McMillan17}.  The corrections applied for the motion of the Sun with respect to the LSR  are $(V_x^{\odot,{\rm pec}},V_y^{\odot,{\rm pec}},V_z^{\odot,\rm pec}) = {\rm (10.0\ km\ s^{-1}, 11.0\ km\ s^{-1}, 7.0\ km\ s^{-1})}$ \citep{Tian15, Bland-Hawthorn16}. The Galactocentric cylindrical components can be calculated:
\begin{align}
& R = \sqrt{x^2+y^2}  \nonumber \\
& \phi =  \tan^{-1}(\frac{y}{x})
\end{align}

\begin{align}
& V_{\phi} = V_x\,\frac{x}{R}+V_y\,\frac{y}{R}  \nonumber \\
& V_R =  -V_x\,\frac{y}{R}+V_y\,\frac{x}{R} \\
& V_z =  V_z, \nonumber
\end{align}
$V_{\phi}$ is in the direction of the Galactic rotation.  Due to the error propagation in the observed quantities,  the uncertainties of the derived parameters for each star are determined by 1,000 realizations of Monte Carlo simulation.  The standard deviation is adopted as uncertainty.
\par
We  integrate the stellar orbits of sample stars based on the observation parameters as the starting point. We use a recent Galactic potential model provided by \cite{McMillan17}. Their model includes five components: the cold gas discs near the Galactic plane, as well as the thin and thick stellar disk, a bulge component, and a dark-matter halo. The GALPOT code \citep{McMillan17,Dehnen98} is used to integrate the stellar orbit and set up orbit integrator with integration time of 1,000 Myr. As a result, we obtain various stellar orbital parameters, such as the closest approach of an orbit to the Galactic center ($ r_{\rm peri}$, i.e., the perigalactic distance),  the farthest extent of an orbit from the Galactic center ($r_{\rm apo}$), the orbital energy ($E$), and angular momentum ($L_{z}$). The orbital eccentricities of sample stars, $e$, defined as $e= (r_{\rm apo} - r_{\rm peri})/(r_{\rm apo} + r_{\rm peri})$.

\subsection{Sample selection}

\par 
\begin{figure*}
	\centering
	\includegraphics[width=1.0\textwidth]{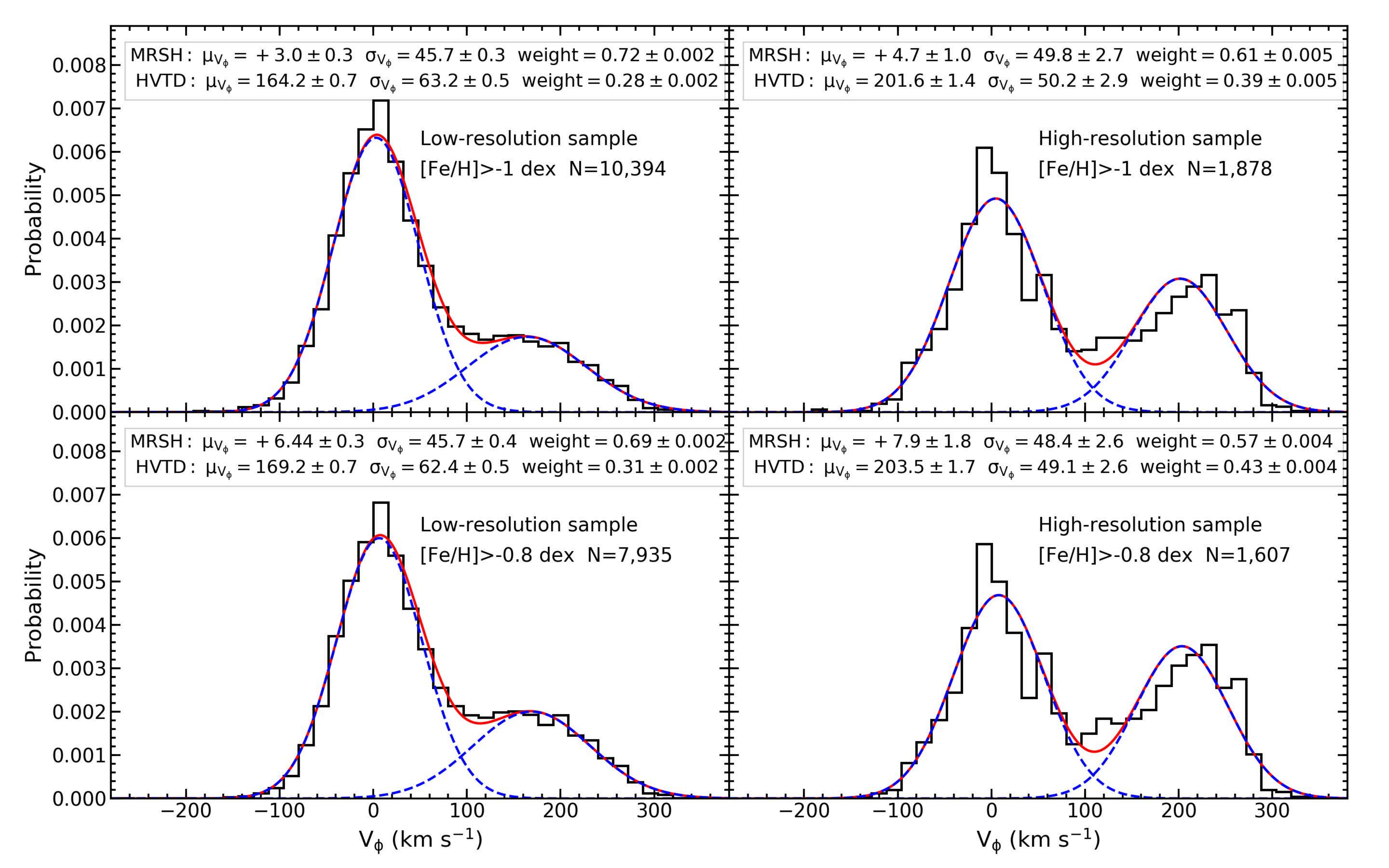}
	\caption{The rotational velocity distribution of the high-velocity sample stars of [Fe/H] $>-1.0$ dex (top panels) and [Fe/H] $>-0.8$ dex (bottom panels). The left and right panels are the low-resolution and high-resolution sample stars, respectively. The distribution functions of the rotational velocity are well fitted with a two-Gaussian model according to the lowest $BIC$. The two single-Gaussian components are interpreted as the metal-rich stellar halo (MRSH) and the high-velocity thick disk (HVTD), and their sum is illustrated by the red curve. The best-fit values of the means ($\mu$), standard deviations ($\sigma$), and weights of each single-Gaussian component  are given in the corresponding panels and $N$  represents the number of stars.}
	\label{figure4}
\end{figure*}
\begin{figure*}
	\centering
	\includegraphics[width=1.0\textwidth]{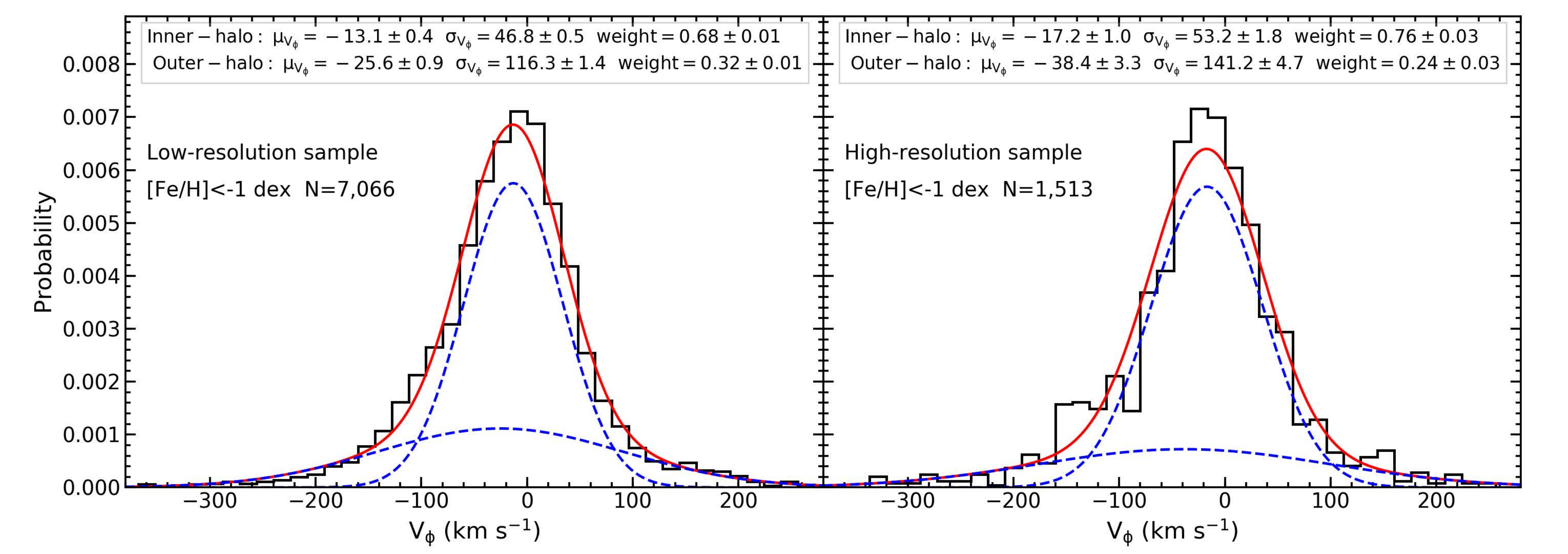}
	\caption{The rotational velocity distribution of the canonical halo stars with $v_{{\rm tot}}>220$ ${\rm km\ s^{-1}}$ and [Fe/H] $<-1.0$ dex for the low-resolution (left panel) and high-resolution sample (right panel). The distribution functions can be fitted with a two-Gaussian model according to the lowest $BIC$. The two single-Gaussian components are regarded as the inner-halo and outer-halo, and their sum is illustrated by the red curve. The best-fit values of the means ($\mu$), standard deviations ($\sigma$), and weights of each single-Gaussian component are given in the corresponding panels.}
	\label{figure5}
\end{figure*}

\par 
The Toomre diagram  has been widely used to  distinguish the thin-disk,
thick-disk, and halo stars, which is a plot of $\sqrt{U_{\rm LSR}^2+W_{\rm LSR}^2}$ versus rotational component $V_{\rm LSR}$.
The halo stars are usually defined as stars with $v_{{\rm tot}} = \sqrt{U_{{\rm LSR} }^2 + V_{{\rm LSR}}^2 + W_{{\rm LSR}}^2 } > 200\sim220$ ${\rm km\ s^{-1}}$ \citep[e.g.,][]{Venn04, Nissen10, Bonaca17,XingZhao18}. In order to investigate the properties of the stellar halo, we define stars with $v_{{\rm tot}} > 220$ ${\rm km\ s^{-1}}$ as our high-velocity sample stars.  According to previous studies, the high-velocity sample stars  mainly consist of halo stars, and their spatial distribution in the Toomre diagram is presented in the top panel of Figure \ref{figure1}. In total,  we obtain 17,470 high-velocity sample stars of  low resolution, and 3,391 of high resolution.  There are 695 common targets between these two samples.
As shown in the bottom panel of Figure \ref{figure1}, our high-velocity sample stars of high-resolution are within $4\lesssim R \lesssim13$ kpc and can extend up to 6 kpc in height from the Galactic plane.

\par 
In this work, the low-resolution sample is used to  study statistically kinematic and chemical characteristics of the stellar halo.
At the same time, since the high-resolution sample has accurate stellar parameters, which is used to confirm the  conclusions derived from the low-resolution sample.  
As shown in Figure \ref{figure2},  the high-resolution sample mainly consists of G- and K-type giant stars, while  the low-resolution sample is mainly A-, F-, G-, and K-type stars.  
Obviously, there are many more low-resolution sample stars than high-resolution sample stars. So low-resolution sample can reduce the influence of sample selection bias.

\section{Kinematics of Metal-rich Stellar Halo and High-Velocity thick disk}
\label{sec:kinematics}
\subsection{Kinematic Evidence of metal-rich stellar halo and high-velocity thick disk}

\par 
Although we have selected halo stars according to kinematic criteria $v_{{\rm tot}}>220$ ${\rm km\ s^{-1}}$,  more high-velocity sample stars are comprised of metal-rich stars ([Fe/H] $>-1.0$ dex) as shown in Figure \ref{figure3}.  
It is similar to the result of \cite{Bonaca17}, who select sample stars within $\lesssim$ 3 kpc from the Sun, based on first Gaia data, the RAVE, and APOGEE spectroscopic surveys.  They regarded these metal-rich stars as metal-rich stellar halo stars.  However, because of these stars exhibit the metallicity of the thick disk, whether these metal-rich stars belong to the halo or disk still need more consideration. Since rotational behavior is a very effective  way to distinguish the thin disk, thick disk, and halo component, we shall further study the rotational velocity distribution. 

\par 
To study how many components these metal-rich stars ([Fe/H] $>-1.0$) contain, 
we first make the traditional assumption that the distribution function of the stellar rotational velocity from a single stellar population is well described by a single-Gaussian function, then the optimal number of the Gaussian function is given by using the Bayesian information criterion ($BIC$) \citep{Ivezic14}:
\begin{align}
BIC=-2ln[L^{0}(M)] + klnN, 
\end{align}
where $L^{0}(M)$ represents the maximum value of the likelihood function of the model, $N$ is the number of data points, and $k$ is the number of free parameters. Uncertainties of the best-fit value  are determined by 1,000 realizations of Monte Carlo simulation and the standard deviations are defined as errors.
Figure \ref{figure4} shows that the rotational velocity distribution of the metal-rich stars can be fitted with a two-peak Gaussian model according to the lowest $BIC$. 
\begin{table*}
	\begin{center}
		\centering
		\caption{ \upshape {The Best-fit Values of Mean ($\mu$), Standard Deviation ($\sigma$), and Weight of Each Rotational Velocity  Distribution of Gaussian Form in Different Metallicity Intervals}}
		\label{Table 1}
		\begin{tabular}{lllllll}
			\hline
			\hline 
			[Fe/H]& \multicolumn{3}{c}{MRSH} & \multicolumn{3}{c}{HVTD}\\
			- & $\mu$  & $\sigma$ & Weight  & $\mu$  & $\sigma$ & Weight \\
			(dex)&(${\rm km\ s^{-1}}$)&(${\rm km\ s^{-1}}$)&-&(${\rm km\ s^{-1}}$)&(${\rm km\ s^{-1}}$)&-\\
			\hline
			\multicolumn{7}{c}{Low Resolution Sample}\\
			\hline
			[Fe/H]$>-1$ & $+3.0\pm0.3$&$45.7\pm0.3$&$0.72\pm0.002$&$164.2\pm0.7$&$63.2\pm0.5$&$0.28\pm0.002$\\
			
			[Fe/H]$>-0.8$ & $+6.44\pm0.3$&$45.7\pm0.4$&$0.69\pm0.002$&$169.2\pm0.7$&$62.4\pm0.5$&$0.31\pm0.002$\\
			\hline
			\multicolumn{7}{c}{High Resolution Sample}\\
			\hline
			[Fe/H]$>-1$ & $+4.7\pm1.0$&$49.8\pm2.7$&$0.61\pm0.005$&$201.6\pm1.4$&$50.2\pm2.9$&$0.39\pm0.005$\\
			
			[Fe/H]$>-0.8$ & $+7.9\pm1.8$&$48.4\pm2.6$&$0.57\pm0.004$&$203.5\pm1.7$&$49.1\pm2.6$&$0.43\pm0.004$\\
			
			\hline
		\end{tabular}
	\end{center}
\end{table*}
\par
Here, we use low and high-resolution samples to confirm each other.  It needs to be noted that the fitted parameters from two samples are slightly different, such as  a best-fit mean rotational velocity is $\mu_{V_{\phi}}=164.2\pm 0.7$ for the low-resolution sample in the top left panel of Figure \ref{figure4}, while $\mu_{V_{\phi}}=201.6\pm 1.4$ for the high-resolution sample  in the top right panel. But we notice that two samples show a consistent component number, which implies that the component number does not depend on the sample. In order to be consistent with the type of high-resolution sample stars,  we also restrict effective temperature with $4000<{\rm T_{eff}}<5300$ K and surface gravity with log(g) $<3.5$ dex for the low-resolution sample.  We find that the  fitted parameters from this restricted low-resolution sample are still slightly different from the high-resolution sample, but the difference between the parameters derived from the high-resolution sample and this restricted low-resolution sample has diminished. So we consider that the differences of parameters from two samples could result from uncertainties of stellar  parameters in the low-resolution sample or incompleteness of the high-resolution sample.

\par
A small number of inner halo stars with [Fe/H] $>-1.0$ dex have been reported by some study \citep[e.g.,][]{An13,An15,Zuo17,Liu18,Gu19}. In order to eliminate  the effects of the inner halo, we also inspect the component number for the stars with [Fe/H] $>-0.8$ dex. Our results show that  the component number is identical for the stars both  [Fe/H] $>-1.0$ and [Fe/H] $>-0.8$ dex as shown in the bottom panel of Figure \ref{figure4}, which indicates the effect of  the inner halo on our results is negligible. Table \ref{Table 1} lists the best-fit values of the two single-Gaussians components in  Figure \ref{figure4}.

\begin{figure*}
	\centering
	\includegraphics[width=1.0\textwidth]{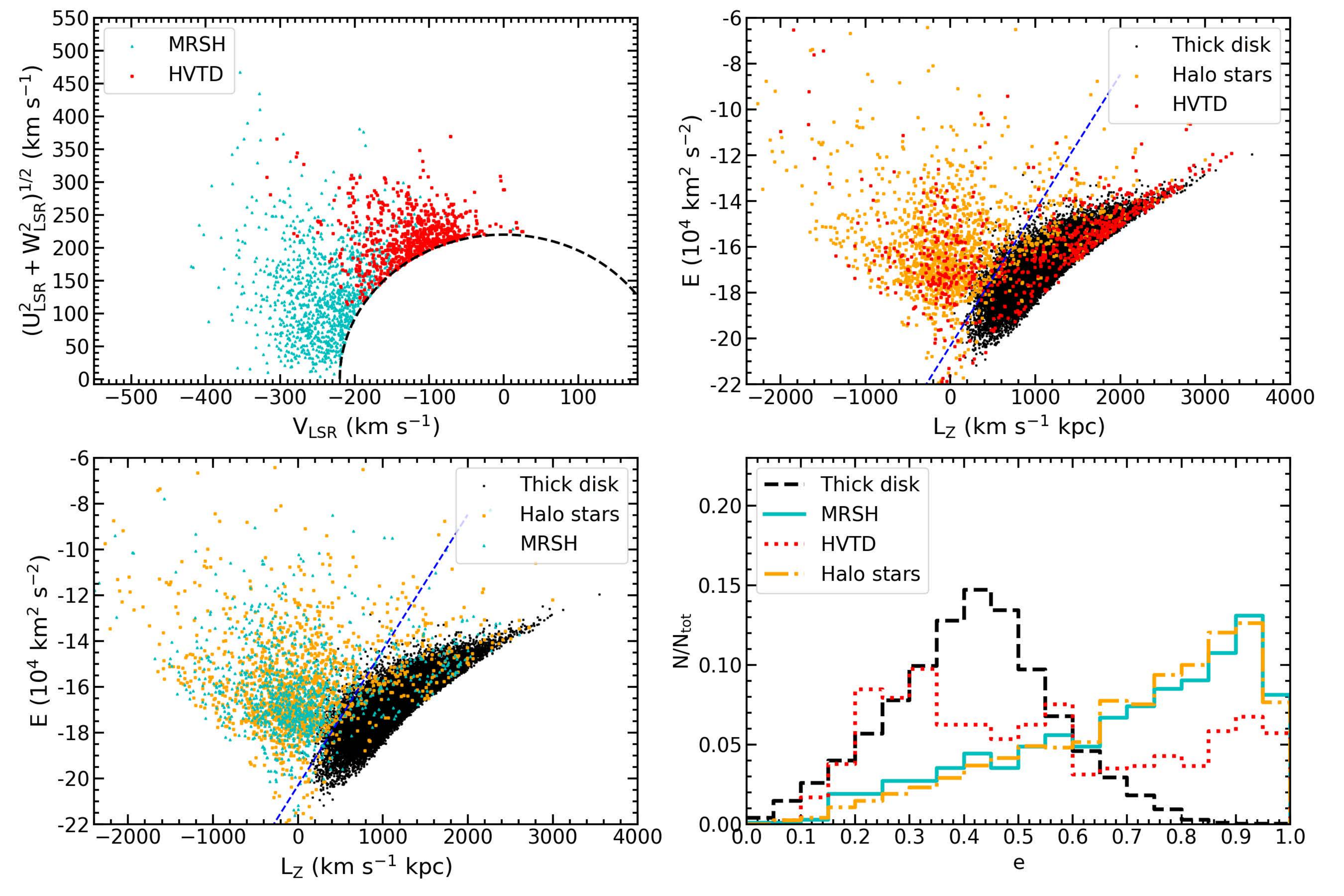}
	\caption{Top left panel: Toomre diagram of the high-velocity thick disk (HVTD, marked by the red dots) and metal-rich stellar halo (MRSH, marked by the cyan dots) for the high-resolution sample. Top right and bottom left panels: The distribution between the total energy and the vertical angular momentum (Lindblad diagram) of the thick disk ( marked by the black dots), HVTD (marked by the red dots), MRSH (marked by the cyan dots), and canonical halo stars (marked by the orange dots).  
		The blue dashed line represents the separation of populations. 
		Bottom right panel: Distribution of orbit eccentricity of the  HVTD (marked by the red dotted line), MRSH (marked by the cyan line), thick disk stars (black dashed line), and canonical halo stars (orange line). }
	\label{figure6}
\end{figure*}
\begin{figure*}
	\centering
	\includegraphics[width=1.0\textwidth]{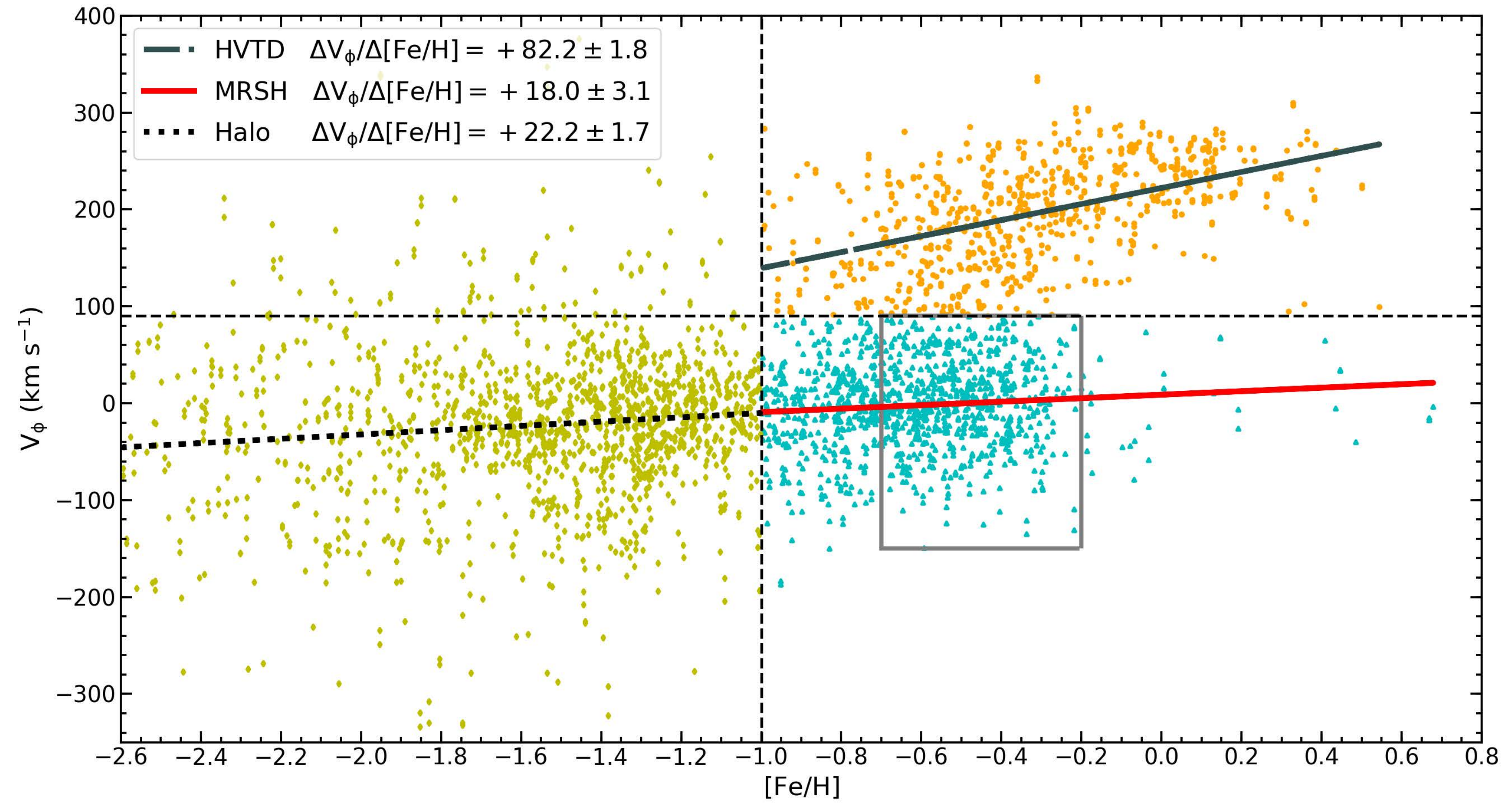}
	\caption{Variation of rotational velocity with metallicity for the MRSH (marked by the cyan dots), HVTD (marked by the orange dots), and halo stars (marked by the yellow-green dots) from our high-resolution sample. These variation trends can be fitted with  linear functions using the least square method. Stars in the gray rectangle box represent Splash stars defined by \cite{Belokurov19}.}
	\label{figure7}
\end{figure*}
\par 
We have confirmed that high-velocity stars of [Fe/H] $>-1.0$ dex contain two independent components using their rotational velocity distribution.  According to previous studies,  the mean rotational velocity of the thick disk is within the range of 160-200 ${\rm km\ s^{-1}}$, and the thin disk is greater than 210 ${\rm km\ s^{-1}}$ \citep[e.g.,][]{Kordopatis11,Li18}. For example, 
\cite{Carollo10} and \cite{Li17,Li18} reported a mean rotational velocity of $<V_{\phi}> \sim 180$ ${\rm km\ s^{-1}}$ for the thick disk, and \cite{Kordopatis11} measured $<V_{\phi}> = 166$ ${\rm km\ s^{-1}}$. 
The halo stars have lower mean rotational velocity \citep[e.g.,][]{Kafle17},  for example, \cite{Smith09} measured $<V_{\phi}> \sim 2.3$ ${\rm km\ s^{-1}}$for the halo stars, \cite{Carollo10} reported $<V_{\phi}> =+7\pm4$ and $<V_{\phi}> =-80\pm13$ ${\rm km\ s^{-1}}$ for the inner halo and outer halo, and \cite{Tian19} reported that local halo have progradely rotates with $<V_{\phi}>\sim +27$. 
As shown in Figure \ref{figure4}, for the stars with [Fe/H]$>-1.0$, one component peaks at  $<V_{\phi}> \sim 201.3$ ${\rm km\ s^{-1}}$ for the high-resolution sample and  $<V_{\phi}> \sim 164.2$ ${\rm km\ s^{-1}}$ for the low-resolution sample, which is consistent with the thick disk. So we consider that this component  should be the high-velocity thick disk (HVTD) and it has the same  rotational velocity and metallicity as the canonical thick disk, but its member stars have the same position as the halo in the Toomre diagram. For the stars with [Fe/H]$>-1.0$, another component peaks at $<V_{\phi}> \sim +4.7$ ${\rm km\ s^{-1}}$ for the high-resolution sample and $<V_{\phi}> \sim +3.0$ ${\rm km\ s^{-1}}$ for the low-resolution sample, which is similar to the  rotational velocity of the halo. Therefore we regard this component as a metal-rich stellar halo (MRSH). It has the same rotational velocity and position as the halo in the Toomre diagram, but it has metallicity of the canonical thick disk. \cite{Belokurov19} measured rotational velocity distribution of the metal-rich stars with $-0.7<$[Fe/H]$<-0.2$ and $2<|z|<3$  on halo-like orbits (Splash stars), with a peak  at  25 ${\rm km\ s^{-1}}$ and standard deviation of $108\pm 19$. They showed that Splash stars contain lots of stars with $V_{\phi} > 100$ ${\rm km\ s^{-1}}$. Because they did not remove disk stars, their Splash stars may be  contaminated by the thin and thick disk stars. 
Since  there exist a  clear  gap between rotational velocity distribution of the HVTD and MRSH as shown in Figure \ref{figure4}, the HVTD is defined as high-velocity sample stars with [Fe/H] $>-1.0$ dex and $V_{\phi} >90$ ${\rm km\ s^{-1}}$, while the MRSH is high-velocity sample stars with [Fe/H] $>-1.0$ dex and $V_{\phi} <90$ ${\rm km\ s^{-1}}$.  
\begin{figure}[]
	\includegraphics[width=1.0\hsize]{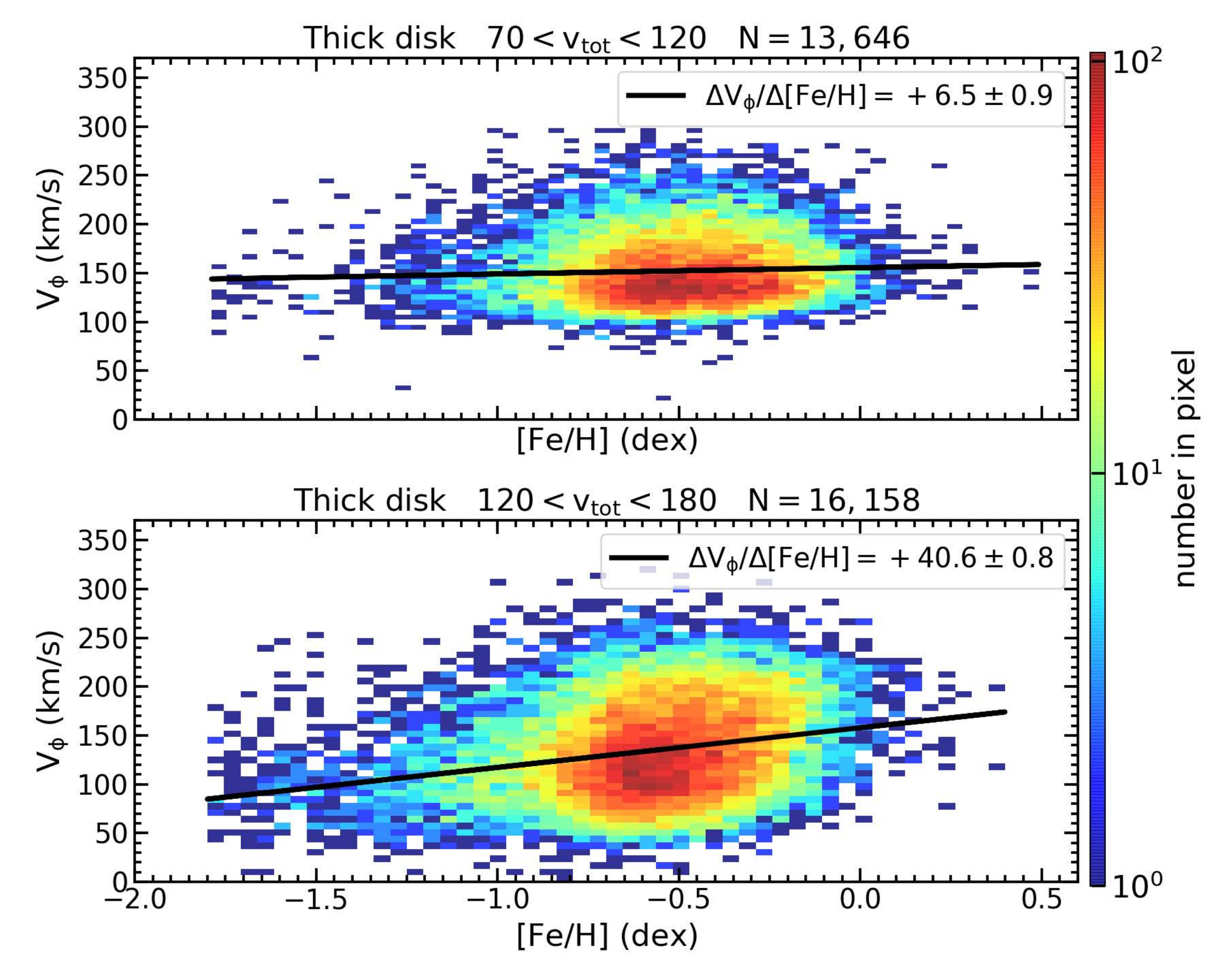}
	\caption{The variation of rotational velocity with metallicity for the thick disk stars of $70<v_{{\rm tot}} <120$ (top panel) and $120<v_{{\rm tot}}<180 $ ${\rm km\ s^{-1}}$ (bottom panel).} 
	\label{figure8}
\end{figure}

\par
We define the canonical halo stars with  $v_{{\rm tot}}>220$ ${\rm km\ s^{-1}}$ and [Fe/H] $<-1.0$ dex.
In order to check whether the canonical halo stars contain  the HVTD or MRSH  stars, we study the rotational velocity distribution of the canonical halo that is fitted with a two-peak Gaussian model according to the lowest $BIC$ as shown in Figure \ref{figure5}.
Some previous studies showed that the canonical stellar halo  has two components: inner-halo and outer-halo. Thus, we could regard these two single-Gauss components as inner-halo and outer-halo. This implies that the stars of $v_{{\rm tot}}>220$ ${\rm km\ s^{-1}}$ and [Fe/H] $<-1$ dex contain very few HVTD or MRSH  stars.     

\begin{figure}
	\includegraphics[width=1.0\hsize]{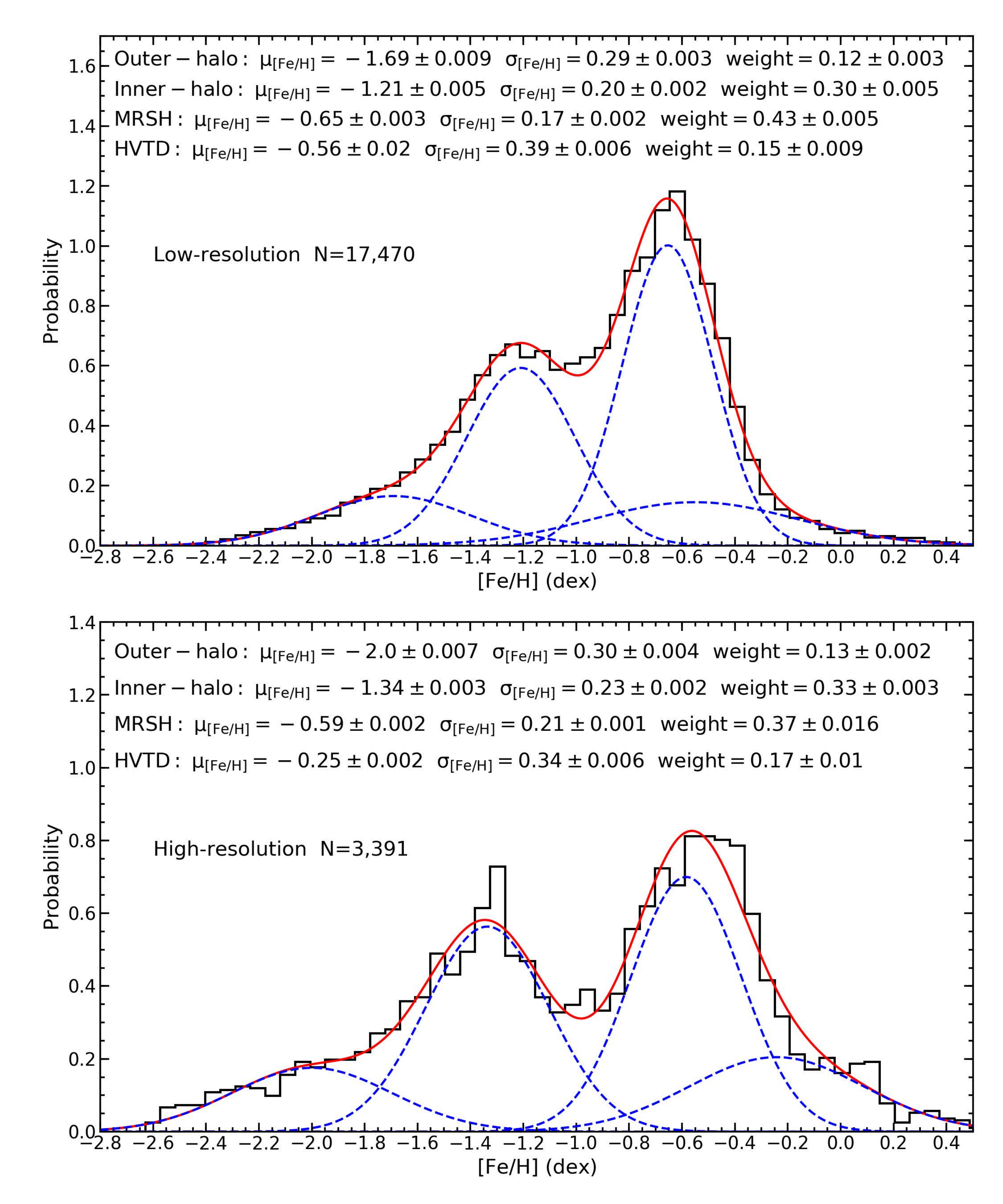}
	\caption{Metallicity distribution of the high-velocity stars for the low-resolution (top panel) and high-resolution sample (bottom panel). The distribution functions are well fitted with a four-Gaussian model according to the lowest $BIC$, which represents the contribution from the outer-halo, inner-halo, MRSH, and HVTD, and the sum is illustrated by the red curve. The best-fit values of the means ($\mu$), standard deviations ($\sigma$), and weights for each single-Gaussian component are given in the corresponding panels. }
	\label{figure9}
\end{figure}
\par 
\begin{figure*}
	\centering
	\includegraphics[width=1.0\textwidth]{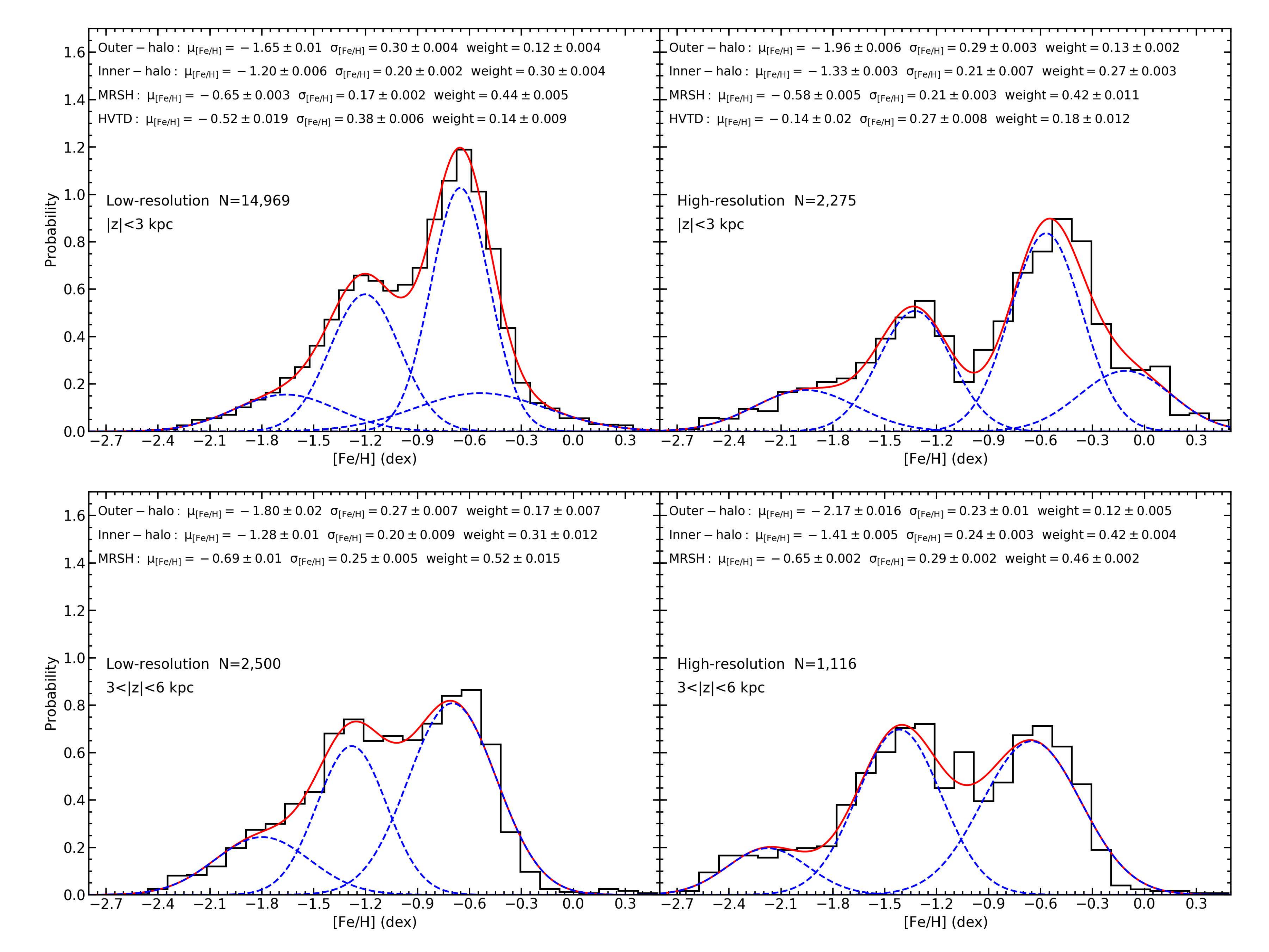}
	\caption{Metallicity distribution of the high-velocity sample stars for the low-resolution (left panel) and high-resolution sample (bottom panel) in different vertical height intervals.  The distribution functions are well fitted with a four-Gaussian or three-Gaussian model according to the lowest $BIC$. The best-fit values of the means ($\mu$), standard deviations ($\sigma$), and weights for each single-Gaussian component are given in the corresponding panels.}
	\label{figure10}
\end{figure*}

\subsection{kinematic properties of the metal-rich stellar halo and high-velocity thick disk}
\par
Figure \ref{figure6} shows the Toomre diagram, Lindblad diagram (a plot of the integrals of motion representing the total energy, $E$, and vertical angular momentum, $L_z$), and distribution of orbit eccentricity for the HVTD, MRSH, thick disk, and canonical halo. Our results indicate that there is a relatively clear separation between the HVTD and  MRSH in the Toomre diagram, but there is a small amount mixing in the boundary. This also implies that  the HVTD and  MRSH could be different population. In addition, in order to present a clear dynamical relation between the HVTD, MRSH, canonical halo, and thick disk, we compare the thick disk sample stars from \cite{Yan19} with the HVTD, MRSH, and canonical halo in Lindblad diagram.
As shown in the top right panel of Figure \ref{figure6}, there is an apparent separation between the canonical halo and thick disk as indicated by the blue dashed line. About $65\%$ of the  HVTD stars  are clustered with the thick disk in the Lindblad diagram, while the other $35\%$ of the HVTD stars are clustered with the canonical halo stars. Furthermore, the  HVTD stars contain $23\%$ of high orbital eccentricity ($e>0.8$) stars as shown in the bottom right panel of Figure \ref{figure6}. After excluding those $35\%$ of the HVTD stars that have the same position as the canonical halo in the Lindblad diagram, the orbit eccentricity distribution of the HVTD is basically consistent with the thick disk. These indicate that our  HVTD stars could be contaminated by the MRSH stars and most of HVTD stars share the same dynamical properties as the thick disk. As shown in the bottom left and right panels of Figure \ref{figure6},  most MRSH stars  are clustered with the canonical halo. The orbit eccentricity distribution of the MRSH is basically consistent with the canonical halo, and most of them have high orbit eccentricity (e$>$0.6), which is  consistent with previous studies \citep[e.g.,][]{Mackereth19b,Fernandez-Alvar19,Belokurov19}. These imply that the MRSH stars share the same dynamical properties as the canonical halo.

\par 
The gradient of rotational velocity with metallicity is important for the Galactic disk, which can provide useful clues to its formation and evolution. Many works have confirmed that the thin disk stars show a negative rotational velocity gradient versus metallicity, and the gradient range from  $\sim$ $-16$ to $-24$ ${\rm km\ s^{-1}\ dex^{-1}}$. The thick disk stars show a positive gradient from  $\sim$ $+30$ to $+49$ ${\rm km\ s^{-1}\ dex^{-1}}$ \citep[e.g.,][]{Lee11,Adibekyan13,Recio-Blanco14,Guiglion15,Jing16,Yan19}.
Figure \ref{figure7} shows the variations of rotational velocity with metallicity for the MRSH, HVTD, and halo stars. Variation trends of the MRSH and HVTD stars can be fitted with linear functions, but their gradients are distinctly different. The HVTD exists a steeper gradient than the canonical thick disk, $\Delta V_{\phi}/\Delta{\rm [Fe/H]} = +82.2 \pm 1.8 $ ${\rm km\ s^{-1}\ dex^{-1}}$. However, the MRSH shows a relatively flat gradient,  $\Delta V_{\phi}/\Delta{\rm [Fe/H]} = +18.0 \pm 3.1 $ ${\rm km\ s^{-1}\ dex^{-1}}$, which is less than the canonical thick disk. The distribution of rotational velocity with metallicity for the halo stars is more scattered than the MRSH and HVTD stars, but it globally exhibits a relatively flat gradient, $\Delta V_{\phi}/\Delta{\rm [Fe/H]} = +22.2 \pm 1.7 $ ${\rm km\ s^{-1}\ dex^{-1}}$. The gradient of the halo is basically equal to that of the MRSH.  \cite{Belokurov19} noticed that some metal-rich stars ([Fe/H]$>-0.7$) show distinct difference  in the rotation velocity versus metallicity distribution for the thin and thick disk stars. The distribution of the rotational velocity with metallicity for these metal-rich stars shows a vertical trend, and these metal-rich stars are referred as Splash stars by \cite{Belokurov19}. The boundaries of the Splash stars by a rectangular box
have been marked in Figure \ref{figure7}. It can be clearly seen that the Splash stars locate in the MRSH.
\par 
The HVTD has a steeper gradient than the canonical thick disk in the rotational velocity versus metallicity distribution, and the gradient of the HVTD is about twice as high as the canonical thick disk. We noticed that the gradient of the rotational velocity versus metallicity in the thick disk depends strongly on the spatial velocity. Figure \ref{figure8} displays variation of rotational velocity with metallicity in different  spatial velocity  intervals for the canonical thick disk from \cite{Yan19}. We can see that the thick disk stars with $70<v_{{\rm tot}}<120$ ${\rm km\ s^{-1}}$ show a very flat gradient, $\Delta V_{\phi}/\Delta{\rm [Fe/H]} = +6.5 \pm 0.9 $ ${\rm km\ s^{-1}\ dex^{-1}}$, while the thick disk stars with $120<v_{{\rm tot}}<180$ ${\rm km\ s^{-1}}$ have a steeper gradient, $\Delta V_{\phi}/\Delta{\rm [Fe/H]} = +40.6 \pm 0.8 $ ${\rm km\ s^{-1}\ dex^{-1}}$.  This implies that the gradient of rotation velocity versus metallicity in the thick disk could increase with the spatial velocity.   In the HVTD stars with $v_{{\rm tot}}>220$ ${\rm km\ s^{-1}}$,  the gradient of rotation velocity  versus metallicity is steeper than the canonical thick disk, which implies that the HVTD stars  could belong to the thick disk.

\section{The Metallicity Distribution of  the metal-rich stellar halo and high-velocity thick disk}
\label{sec:metallicity}
\par 

We  obtained the HVTD or MRSH stars selected by rotational velocity distribution and metallicity.
The mean metallicities of the  HVTD stars ${\rm <[Fe/H]>=-0.51 \pm 0.002}$ dex with standard deviations of  $\sigma_{\rm [Fe/H]}=0.26$  dex 
for the low-resolution sample, and ${\rm <[Fe/H]>=-0.31 \pm 0.0004}$ dex with  standard deviations $\sigma_{\rm [Fe/H]}=0.31 $ dex
for the high-resolution sample.  The mean metallicities of the  MRSH stars are ${\rm <[Fe/H]>=-0.67 \pm 0.001}$ dex with standard deviations $\sigma_{\rm [Fe/H]}=0.20 $ dex 
for the low-resolution sample and ${\rm <[Fe/H]>=-0.60 \pm 0.0003}$ dex with standard deviations $\sigma_{\rm [Fe/H]}=0.23$ dex for the high-resolution sample. Therefore, the HVTD stars have higher metallicity than the MRSH  on average.
The metallicity distributions of both high-velocity sample stars are well fitted with a four-peak Gaussian model according to the lowest $BIC$ in Figure \ref{figure9}. 
The two single-Gaussian for the canonical halo stars with [Fe/H] $\lesssim-1$ dex could be interpreted as the inner-halo and outer-halo.  
The relative metal-rich stars with [Fe/H] $\gtrsim-1$ dex also exist two single-Gausses components, which could be interpreted as the  HVTD and MRSH stars. As shown in Figure \ref{figure9}, the canonical halo contains few HVTD and MRSH.

\begin{table}
	\begin{center}
		\centering
		\caption{ \upshape {The Best-fit Values of Mean ($\mu$), Standard Deviation ($\sigma$), and Weight (W) of Each  Metallicity  Distribution of Gaussian Form in Different Vertical Height Intervals.} }
		\label{Table 2}
		\begin{tabular}{lllll}
			\hline
			\hline 
			& Outer-halo&Inner-halo &MRSH&HVTD\\
			\hline
			\multicolumn{5}{c}{Low Resolution Sample   $|z|\lesssim6$ kpc}\\
			\hline
			$\mu$&$-1.69\pm0.009$&$-1.21\pm0.005$&$-0.65\pm0.003$&$-0.56\pm0.02$\\
			$\sigma$ &$0.29\pm0.003$&$0.20\pm0.002$&$0.17\pm0.002$&$0.39\pm0.006$\\
			W&$0.12\pm0.003$&$0.30\pm0.005$&$0.43\pm0.005$&$0.15\pm0.009$\\
			\hline
			\multicolumn{5}{c}{High Resolution Sample   $|z|\lesssim6$ kpc}\\
			\hline
			$\mu$ &$-2.0\pm0.007$&$-1.34\pm0.003$&$-0.59\pm0.002$&$-0.25\pm0.002$\\
			$\sigma$ &$0.30\pm0.004$&$0.23\pm0.002$&$0.21\pm0.001$&$0.34\pm0.006$\\
			W&$0.13\pm0.002$&$0.33\pm0.003$&$0.37\pm0.016$&$0.17\pm0.01$\\
			
			\hline
			\multicolumn{5}{c}{Low Resolution Sample   $|z|<3$ kpc}\\
			\hline
			$\mu$ &$-1.65\pm0.01$&$-1.20\pm0.006$&$-0.65\pm0.003$&$-0.53\pm0.019$\\
			$\sigma$ &$0.30\pm0.004$&$0.20\pm0.002$&$0.17\pm0.002$&$0.38\pm0.006$\\
			W&$0.12\pm0.004$&$0.30\pm0.004$&$0.44\pm0.005$&$0.14\pm0.009$\\
			\hline
			\multicolumn{5}{c}{High Resolution Sample   $|z|<3$ kpc}\\
			\hline
			$\mu$ &$-1.96\pm0.006$&$-1.33\pm0.003$&$-0.58\pm0.005$&$-0.14\pm0.02$\\
			$\sigma$ &$0.29\pm0.003$&$0.21\pm0.007$&$0.21\pm0.003$&$0.27\pm0.008$\\
			W&$0.13\pm0.002$&$0.27\pm0.003$&$0.42\pm0.011$&$0.18\pm0.012$\\
			
			\hline
			\multicolumn{5}{c}{Low Resolution Sample   $3<|z|<6$ kpc}\\
			\hline
			$\mu$ &$-1.80\pm0.02$&$-1.28\pm0.01$&$-0.69\pm0.01$&-\\
			$\sigma$&$0.27\pm0.007$&$0.20\pm0.009$&$0.25\pm0.005$&-\\
			W&$0.17\pm0.007$&$0.31\pm0.012$&$0.52\pm0.015$&-\\
			\hline
			\multicolumn{5}{c}{High Resolution Sample   $3<|z|<6$ kpc}\\
			\hline
			$\mu$ &$-2.17\pm0.016$&$-1.41\pm0.005$&$-0.65\pm0.002$&-\\
			$\sigma$&$0.23\pm0.01$&$0.24\pm0.003$&$0.29\pm0.002$&-\\
			W&$0.12\pm0.005$&$0.42\pm0.004$&$0.46\pm0.002$&-\\
			\hline
		\end{tabular}
	\end{center}
\end{table}
\par

We noticed that the parameters fitted by the low-resolution and high-resolution sample are slightly different.   But two samples show consistent component number, which implies that the component number does not depend on the sample.   In addition, we restrict the effective temperature with $4000<{\rm T_{eff}}<5300$ K and surface gravity with log(g) $<3.5$ dex for the low-resolution sample in order to be consistent with the high-resolution sample stars.  We find that the  parameters fitted by this restricted low-resolution sample are still slightly different from the high-resolution sample but the difference has diminished.   So the parameter differences from two samples could result from the metallicity uncertainty of the low-resolution sample.

\par 
Thus, we have also confirmed the existence of the HVTD and MRSH  by the metallicity distribution.  
We now study the variation of the MDFs  with vertical distance for these high-velocity sample stars.
Figure \ref{figure10} shows the lowest $BIC$ fitting  from the data: a four-peak and three-peak Gaussian model in the low- and high-resolution sample of $|z|<3$ kpc and $3<|z|<6$ kpc, respectively.  The top panel of Figure \ref{figure10} shows that there are four components within  $|z|<3$ kpc: outer-halo, inner-halo, MRSH, and HVTD. The inner-halo and MRSH occupy the vast majority, and the outer-halo component  still exists within $|z|<3$ kpc. The bottom panel of  Figure \ref{figure10} shows that there are three components in  $|z|>3$ kpc: outer-halo, inner-halo, and MRSH. The inner-halo and MRSH still occupy the majority, but their weights are higher than that of $|z|<3$ kpc. 
The  weight of the outer-halo component is basically invariable, which implies that vertical height has little effect on the outer-halo component within $|z|<6$ kpc. Furthermore, when $|z|>3$ kpc, the HVTD component disappeared, which indicates that most of the HVTD stars are within  $|z|<3$ kpc. Therefore, the variation of component weight with vertical height also indicates that the MRSH stars belong to the halo, and HVTD stars attribute to the thick disk. Table \ref{Table 2} lists the best-fit values of the four or three single-Gaussian components in Figure \ref{figure9} and Figure \ref{figure10}.  \cite{Belokurov19} used K-giants identified in the Sloan Digital Sky Survey spectroscopy to show that the Splash population extends as far as $|z|\sim 20$ kpc, and the ranking of the vertical  sizes of the Splash, the disc and the halo, i.e. $z_{\rm disc}<z_{\rm Splash}<z_{\rm halo}$, which are consistent with the result of our MRSH. 

\par 
The existence of metal-weak thick disk (MWTD) has been confirmed by several works, such as \cite{Morrison90}, \cite{Beers95}, \cite{Chiba00} and \cite{Beers02,Beers14}. The MWTD has disk-like kinematics \citep[e.g.,][]{Chiba00, Carollo10}, and is a low-metallicity tail of the thick disk \citep[e.g.,][]{Morrison90, Beers14, Yan19}. \cite{Ivezic08} and \cite{Carollo10} tried to prove the MWTD as an independent stellar population from the thick disk and revealed that its mean rotational velocity could in the range of 100-150 ${\rm km\ s^{-1}}$, while its metallicity values span from $-0.8$ dex to $-1.7$ dex. \cite{ Kordopatis13} reported that its lowest metallicity at least go down to [M/H] $=-2.0$ dex. Recently, \cite{Carollo19} reported that the MWTD contains two-times less metal content than the canonical thick disk and exhibits enrichment of light elements typical of the oldest stellar populations of the Galaxy, and its rotational velocity is $\sim 150$ ${\rm km\ s^{-1}}$, with a velocity dispersion $60$ ${\rm km\ s^{-1}}$. These properties of the MWTD, including its velocity components and metallicity range, are different from the HVTD and MRSH. Therefore, we consider that the HVTD and MRSH could be two different stellar populations from the MWTD.

\section{Discussion on the Potential Origins of the MRSH and HVTD}

\label{sec:discussion}
\begin{figure*}
	\centering
	\includegraphics[width=1.0\textwidth]{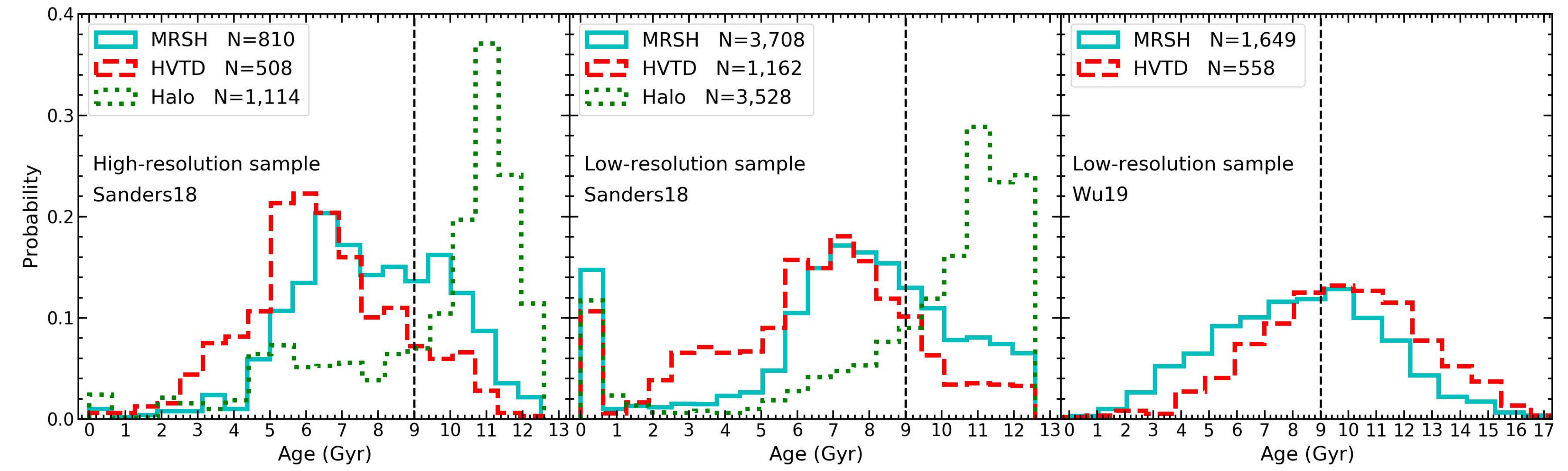}
	\caption{Age distributions of the MRSH, HVTD, and halo stars. The ages of the stars in the left (middle) panel is obtained by cross-matching between our high-resolution (low-resolution) sample and \cite{Sanders18} catalog. The ages of stars in the right panel are obtained by cross-matching between our low-resolution and \cite{Wu19} catalog. $N$ represents the number of stars. The black dashed lines represent the epoch of the massive merger event (Gaia-Sausage).}
	\label{figure11}
\end{figure*}
\par 
Some previous works have suggested that the metal-rich stars with thick-disk metallicity  on halo-like orbits have likely been born in-situ rather than having been accreted from satellite systems \citep[e.g.,][]{Bonaca17,Haywood18,Di Matteo18,Gallart19,Belokurov19}.  \cite{Bonaca17} kinematically identified halo stars in the solar neighborhood with relative speeds larger than 220 ${\rm km\ s^{-1}}$ with respect to the local standard of rest based on the RAVE and APOGEE spectroscopic surveys. Because the orbital directions of the metal-rich stars with [Fe/H] $>-1.0$ are preferentially aligned with the disk rotation, they proposed that these metal-rich halo stars may have formed in situ, rather than having been accreted from satellite systems, and these metal-rich halo stars have likely undergone substantial radial migration or heating. In addition, as a part of the metal-rich halo stars, the Splash stars  have chemical and kinematic properties similar to our MRSH stars. Because the Splash stars are predominantly old, but not so old as the stars deposited into the Milky Way in the last major merger, \cite{Belokurov19} concluded that the Splash stars may have been born in the Milky Way’s proto-disc prior to the massive ancient accretion event which drastically altered their orbits, and they put constraints of the epoch of the last massive accretion event to have finished 9.5 Gyr ago. This massive ancient merger event is the Gaia-Sausage \citep{Belokurov18,Myeong18} (sometimes also referred as Gaia-Enceladus \citep{Helmi18}). Therefore, according to the chemical and kinematic properties, it implies that the MRSH stars were born in situ and the HVTD stars are a part of the thick disk. 

\par 
On the other hand, the stellar ages are also an effective  way to  probe the potential origins of the population. However, it is difficult to obtain accurate stellar ages, and different methods of estimating age  have systematic differences \citep{Frankel19}. In this work, we only use the age range of the stars to discuss the  potential origins of the MRSH and HVTD. Because the Gaia-Sausage merger could happen $\sim 9-11$ Gyr ago \citep[e.g.,][]{Belokurov18,Helmi18,Di Matteo18,Belokurov19}, we define old stars as older than $9$ Gyr and young stars as younger than $9$ Gyr. Ages of our sample stars are obtained by cross-matching with two catalogs, Sanders18 catalog \citep{Sanders18} and Wu19 catalog \citep{Wu19}.  
\cite{Sanders18} presented a catalog of stellar  distances, masses, and ages for $\sim3$ million giant  stars.  The mass and ages have been estimated using the method outlined in \cite{Das19}.  \cite{Sanders18} only estimated masses and ages for the stars metal-richer than $-1.5$ dex  and the maximum age isochrone considered is 12.6 Gyr.  
\cite{Wu19} presented a catalog of stellar age and mass estimates for red giant branches (RGB) stars from the LAMOST DR4. The estimated age has a median error of 30 percent for the stars of SNR $>30$.
The age distributions of the MRSH and HVTD stars are shown in Figure \ref{figure11}.  Although the age distributions of the MRSH and HVTD stars inferred from different samples and methods have  some differences, these age distributions confirm that both MRSH and HVTD stars contain a certain number of young stars ($<9$ Gyr) and old stars ($>9$ Gyr). 
\par
For the young stars ($<9$ Gyr),  their formation may not be affected by the Gaia-Sausage merger. In this regard, the MRSH stars were likely born in-situ rather than accreted from the Gaia-Sausage merger.  The in-situ population can contain stars formed in the initial gas collapse \citep{Samland03} and/or stars formed in the disk,
which has subsequently been kicked out and placed on halo \citep{Zolotov09,Purcell10}. However, it is difficult to distinguish  between the MRSH stars formed by the initial gas collapse and  being heated from the disk. \cite{Cooper15} listed two different channels that the initial gas collapse to form the in-situ stellar halo: stars formed from gas smoothly accreted on to the halo and stars formed in streams of gas stripped from infalling satellites.  The `phase wrapping' signature  in the disk \citep[e.g.,][]{Fux01,Minchev09,Gomez12,de la Vega15} and some substructures in the phase space, such as the Gaia snail and spiral \citep{Antoja18}, are now widely considered to be a relic of a recent external perturbation by a satellite or dwarf galaxy flyby  such as Sagittarius \citep[e.g.,][]{Antoja18,Binney18,Laporte19,Bland-Hawthorn19}. In particular, the last pericentre of the orbit of the Sagittarius has been shown to have a strong effect on the disk stars \citep{Purcell10,Gomez12,de la Vega15}. These external perturbations may heat up disk stars \citep{Mackereth19a} and subsequently, alter their orbits. In addition, the radial migration may also explain the origins of the MRSH stars. \cite{El-Badry16} reported that stars in low-mass galaxies experience significant radial migration via two related processes. First, inflowing and outflowing gas clouds are driven by stellar feedback can remain star-forming, and initial orbits of producing stars can be eccentric and have large anisotropic. Second, outflows and inflows gas drive strong fluctuations in the overall galactic potential, and stellar orbits are affected by such fluctuations, ultimately becoming heated. \cite{Bonaca17} concluded that this radial migration mechanism could explain the origin of metal-rich stars on halo-like orbits in the solar neighborhood. 

\par 
For the old stars formed in-situ ($>9$ Gyr), the Gaia-Sausage merger event may have a major effect on their formation. The MRSH stars may form in an old proto-disk, possibly dynamically heated by the Gaia-Sausage merger, and subsequently be kicked out to the halo. This result is  consistent with previous studies \citep[e.g.,][]{Haywood18,Di Matteo18,Gallart19,Belokurov19}. The HVTD stars also form in an old proto-disk, but these stars may be so less affected by the Gaia-Sausage merger event than MRSH that they retain some properties of the thick disk.

\section{Summary and Conclusions}
\label{sec:summary}
\par 
Based on a  high-resolution sample of  G-/K-type giant stars from the APOGEE DR14 
and a low-resolution sample of A-, F-, G-, and K-type stars from the LAMOST spectroscopic survey combined Gaia DR2 survey, 
we obtained high-velocity sample stars ($v_{{\rm tot}}>220$ ${\rm km\ s^{-1}}$) in the Toomre diagram. 
From the kinematic and chemical distribution of these high-velocity sample stars, we  concluded that the Galaxy exists 
a metal-rich stellar halo (MRSH) and a high-velocity thick disk (HVTD), and studied their kinematic and chemical properties. 

\par
The rotational velocity distribution of the sample  stars with $v_{{\rm tot}}>220$ ${\rm km\ s^{-1}}$ and [Fe/H] $>-1.0$ dex can be well described by a two-Gaussian model, associated with the HVTD and MRSH. We also confirmed that the metallicity distribution of the sample stars can be described by a four-Gaussian model: outer-halo, inner-halo, MRSH, and HVTD.   The HVTD  has  basically the same  rotational velocity and metallicity as the canonical thick disk, and it shares the same dynamical properties as the thick disk. However, their member stars have the same position as the halo in the Toomre diagram. The MRSH shows basically the  same  rotational velocity, orbit eccentricity, and position in the Lindblad and Toomre diagram as the canonical halo, but their metallicity distribution similar to the thick disk.   

\par
In addition, we found that the outer-halo component  still exists within $|z|<3$ kpc, and the increase of vertical height has little effect on the proportions of the outer-halo component in $|z|<6$ kpc. Among these stars with  $v_{{\rm tot}}>220$ ${\rm km\ s^{-1}}$, $|z|<6$ kpc, and $4\lesssim R \lesssim13$ kpc, the inner-halo, and MRSH occupy the vast majority, and most of the HVTD stars are within  $|z|<3$ kpc and they have higher metallicity than the  MRSH  stars on average, and the canonical halo contains very few HVTD or MRSH  stars. 
For the HVTD, there exist a steeper gradient of rotational velocity with metallicity than canonical thick disk.  However, the gradient of rotational velocity with metallicity for the MRSH is  more flat than the canonical thick disk. Their chemical and kinematic properties and age imply that the MRSH and HVTD stars may form in situ rather than being accreted from satellite systems.

\section{Acknowledgements}
\par 
We thank especially the referee for insightful comments and suggestions, which has improved the paper significantly.
This work was supported by the National Natural Foundation of China (NSFC No. 11973042 and No. 11973052).
This project was developed in part at the 2016 NYC Gaia Sprint, hosted by the Center for Computational Astrophysics at the Simons Foundation in New York City. The Guoshoujing Telescope (the Large Sky Area Multi-Object Fiber Spectroscopic Telescope, LAMOST) is a National Major Scientific Project built by the Chinese Academy of Sciences. Funding for the project has been provided by the National Development and Reform Commission. LAMOST is operated and managed by the National Astronomical Observatories, Chinese Academy of Sciences. 
\par 
Funding for the Sloan Digital Sky Survey IV has been provided by the Alfred P. Sloan Foundation, the U.S. Department of Energy Office of Science, and the Participating Institutions. SDSS-IV acknowledges
support and resources from the Center for High-Performance Computing at
the University of Utah. The SDSS web site is www.sdss.org.
\par 
SDSS-IV is managed by the Astrophysical Research Consortium for the
Participating Institutions of the SDSS Collaboration including the
Brazilian Participation Group, the Carnegie Institution for Science,
Carnegie Mellon University, the Chilean Participation Group, the French Participation Group, Harvard-Smithsonian Center for Astrophysics,
Instituto de Astrof\'isica de Canarias, Johns Hopkins University,
Kavli Institute for the Physics and Mathematics of the Universe (IPMU) /
University of Tokyo, Lawrence Berkeley National Laboratory,
Leibniz Institut f\"ur Astrophysik Potsdam (AIP),
Max-Planck-Institut f\"ur Astronomie (MPIA Heidelberg),
Max-Planck-Institut f\"ur Astrophysik (MPA Garching),
Max-Planck-Institut f\"ur Extraterrestrische Physik (MPE),
National Astronomical Observatories of China, New Mexico State University,
New York University, University of Notre Dame,
Observat\'ario Nacional / MCTI, The Ohio State University,
Pennsylvania State University, Shanghai Astronomical Observatory,
United Kingdom Participation Group,
Universidad Nacional Aut\'onoma de M\'exico, University of Arizona,
University of Colorado Boulder, University of Oxford, University of Portsmouth,
University of Utah, University of Virginia, University of Washington, University of Wisconsin,
Vanderbilt University, and Yale University.
\par 
This work has made use of data from the European Space Agency (ESA) mission Gaia (http://www.cosmos.esa.int/gaia), processed by the Gaia Data Processing and Analysis Consortium (DPAC, http://www.cosmos.esa.int/web/gaia/dpac/consortium). Funding for DPAC has been provided by national institutions, in particular the institutions participating in the Gaia Multilateral Agreement.

\appendix  
  
\section{Distance and velocity estimation}
\label{appendix}
\par 
We determined the distance and velocity for more than 300,000 giant stars from the LSS-GAC DR4 and Gaia DR2 catalog \citep{Yan19}. Here, we use the Bayesian approach to estimate the distance and velocity of the sample stars with $\sigma_{\varpi}/(\varpi-\varpi_{\rm zp}) \geq 0.1$  \citep{Bailer-Jones15,Bailer-Jones18, Astraatmadja16a, Astraatmadja16b, Luri18}, and compare the results with other works. 
\par
Our goal is to obtain the posterior probability $ P(\bm{{\rm \theta}}|\bm{{\rm x}})$ of observed stars, and the posterior can be written as :
\begin{eqnarray}
P(\bm{{\rm \theta}}|\bm{{\rm x}})\propto P(\bm{{\rm x}}|\bm{{\rm \theta}})P(\bm{{\rm \theta}})= \textrm{exp}[-\frac{1}{2}(\bm{{\rm x}}-\bm{{\rm m}}(\bm{{\rm \theta}}))^{\rm T}\Sigma^{-1}(\bm{{\rm x}}-\bm{{\rm m}}(\bm{{\rm \theta}}))]P(\bm{{\rm \theta}}),
\end{eqnarray}
where data vector $\bm{{\rm x}} = (\varpi, \mu_{\alpha^*}, \mu_{\delta})^{\rm T}$, $\varpi$ denotes stellar parallax, $\mu_{\alpha^*}$ and $\mu_{\delta}$ are proper motion in right ascension and declination respectively, and the symbol `T' represents matrix transpose. The parameters vector is written as $\bm{{\rm \theta}} = (d, v, \phi)^{\rm T}$, consist of the heliocentric distance ($d$), tangential speed ($v$), and travel direction ($ \phi$, increasing anti-clockwise from North). $ \bm{{\rm m}}= (\frac{10^3}{d}, {c_2} \frac{10^3v  \sin\phi}{d}, {c_2} \frac{10^3v  \cos\phi}{d})^{\rm T}$, where $c_2 = ({\rm pc\cdot mas\cdot yr^{-1} })/({\rm 4.74\cdot km\ s^{-1}})$. ${\rm \Sigma}$ is a covariance matrix:

\begin{align}
{\rm \Sigma}=
\begin{pmatrix} \sigma^2_{\varpi} & \sigma_{\varpi}\sigma_{ \mu_{\alpha^*}}\rho(\varpi,\mu_{\alpha^*}) & \sigma_{\varpi}\sigma_{\mu_{\delta}}\rho(\varpi,\mu_{\delta}) \\ \sigma_{\varpi}\sigma_{ \mu_{\alpha^*}}\rho(\varpi,\mu_{\alpha^*}) & \sigma^2_{\mu_{\alpha^*}} & \sigma_{ \mu_{\alpha^*}}\sigma_{\mu_{\delta}}\rho(\mu_{\alpha^*},\mu_{\delta}) \\ \sigma_{\varpi}\sigma_{\mu_{\delta}}\rho(\varpi,\mu_{\delta}) & \sigma_{ \mu_{\alpha^*}}\sigma_{\mu_{\delta}}\rho(\mu_{\alpha^*},\mu_{\delta}) & \sigma_{\mu_{\delta}}^2  \end{pmatrix},
\end{align}
where $\rho(i,j)$ denotes the correlation coefficient between the astrometric parameters $i$ and $j$. $\sigma_k$ is the error of astrometric parameters $k$. $P(\bm{{\rm \theta}})$ represents the prior distribution of the parameters vector, $P(\bm{{\rm \theta}})=P(d)P(v)P(\phi) $\citep{Luri18}, here 
\begin{align}
P(d) &\propto \begin{cases} d^2e^{-d/L(a,b)}& d>0\\ 0& d\le 0 \end{cases}\\
P(v) &\propto \begin{cases} (\frac{v}{v_{\rm max}})^{\alpha-1}(1-\frac{v}{v_{\rm max}})^{\beta-1}& {\rm if}\ 0 \le v \le v_{\rm max}\\ 0& \text{otherwise} \end{cases}\\
P(\phi) &\propto \frac{1}{2\pi}.
\end{align}
We use the exponentially decreasing space density prior for distance\citep{Bailer-Jones15,Bailer-Jones18}, and  adopt the length scale of the Galactic longitude and latitude dependent \citep{Bailer-Jones18}, $L(a,b)$, which is obtained by fitting a spherical harmonic model. We assume that the prior over the angle $\phi$ is uniform. The prior over speed is a beta distribution, and we adopt $\alpha=2$, $\beta=3$ and $ v_{\rm max}=750\ {\rm km\ s^{-1}}$.
\begin{figure*}
	\centering
	\includegraphics[width=1.0\textwidth]{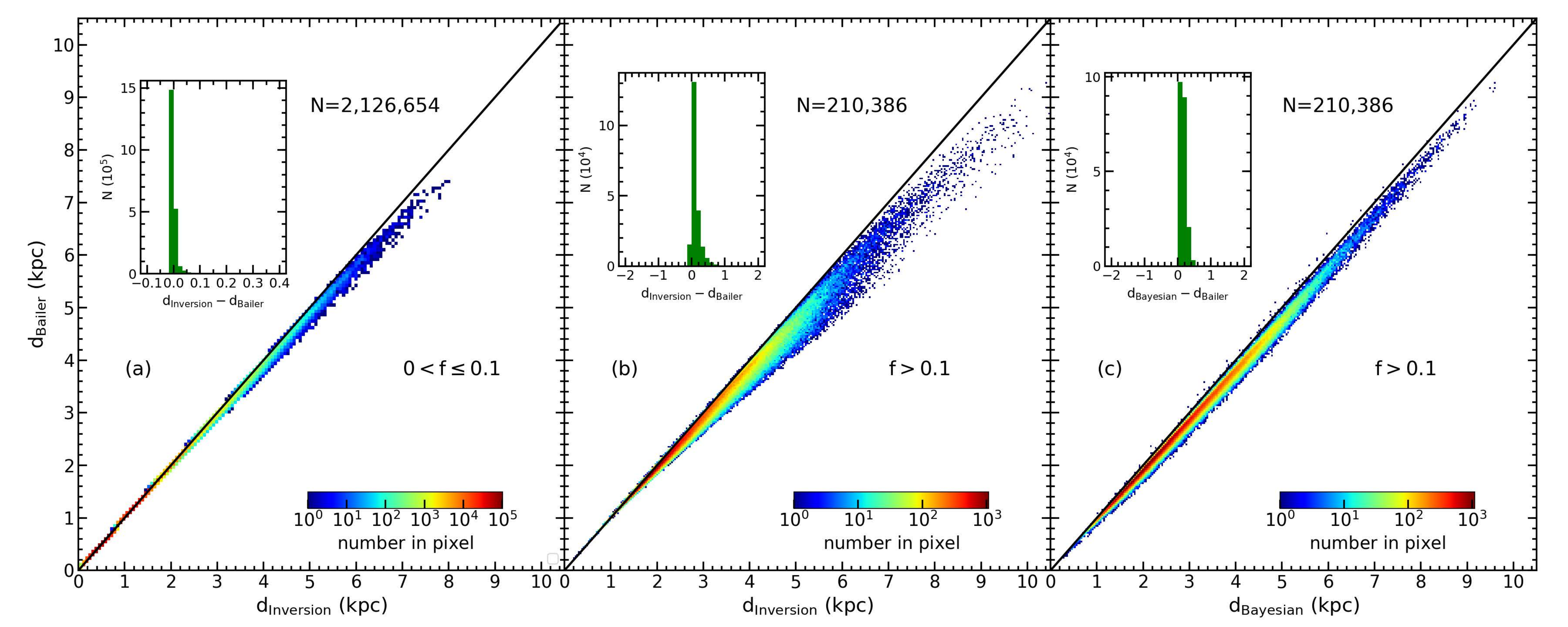}
	\caption{Comparisons of the distance estimation of our sample stars  with \cite{Bailer-Jones18} ($d_{\rm Bailer}$). The stars in  panel (a)  have a relative error in parallax $0<f \equiv \sigma_{\varpi}/(\varpi-\varpi_{\rm zp}) \leq 0.1$. Their distances ($d_{\rm Inversion}$) are obtained by inverting the parallax. The stars in panel (b) and (c) have a relative error in parallax $f>0.1$. The distances of the former ($d_{\rm Inversion}$) are calculated by inverting the parallax, while the distances of the latter ($d_{\rm Bayesian}$) is determined using Bayesian analysis. The black solid line represents the $1:1$ line. $N$ represents the number of subsample stars.}
	\label{figure_app0}
\end{figure*}

\par 
\begin{figure*}
	\centering
	\includegraphics[width=1.0\textwidth]{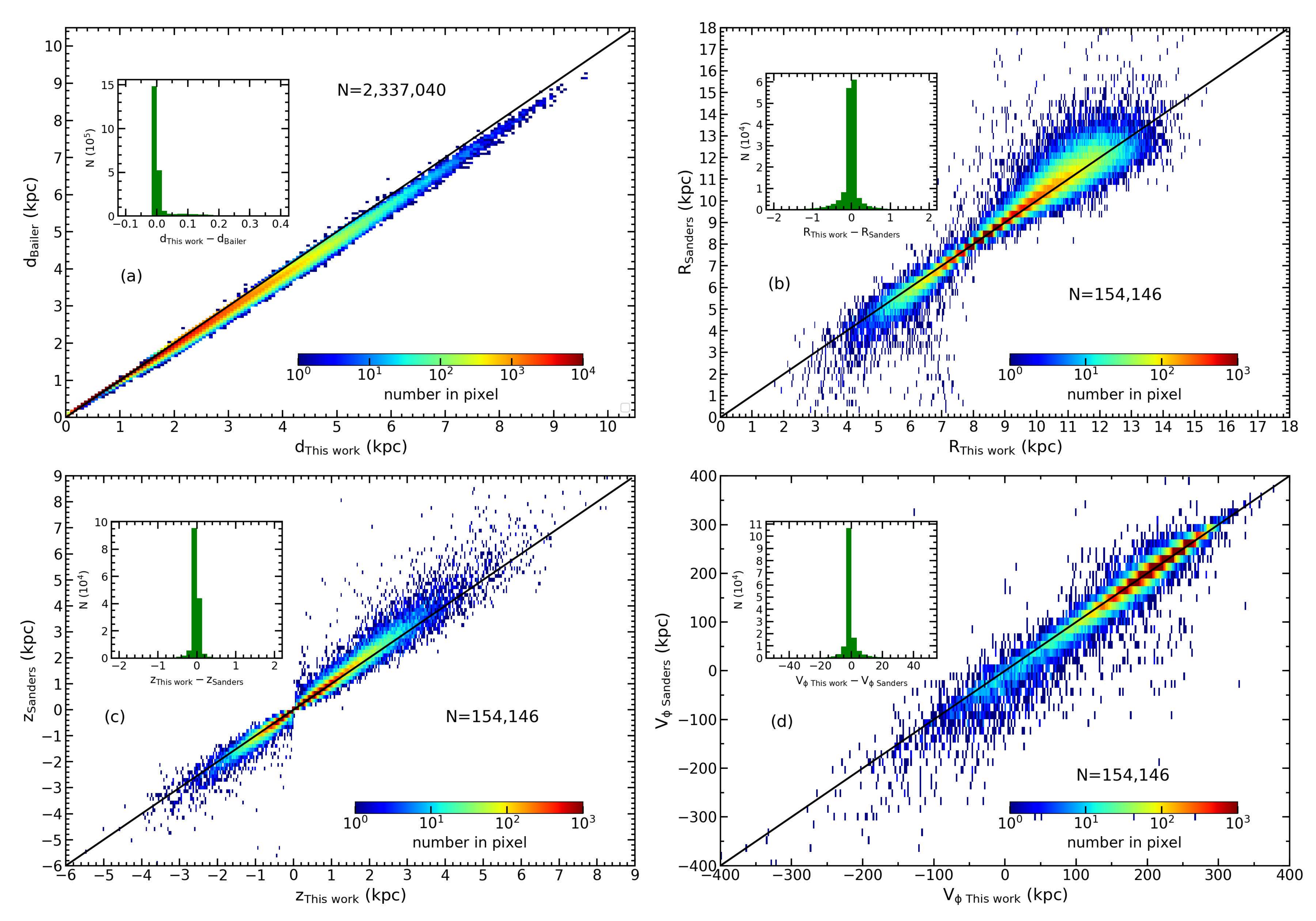}
	\caption{ Distance and velocity estimation comparisons. Panel (a) compares the distance of this work with Bailer-Jones18 catalog \citep{Bailer-Jones18}. Panel (b), (c), and (d) compare the radial distance, vertical height, and rotational velocity in Galactocentric cylindrical coordinate with Sanders18 catalog \citep{Sanders18}. The color bar represents the number of stars. The black solid line represents the $1:1$ line.}
	\label{figure_app1}
\end{figure*}
The posterior probability $ P(\bm{{\rm \theta}}|\bm{{\rm x}})$ of observed stars can be obtained by using  the formulas (A1). We use Markov chain Monte Carlo (MCMC) sampler EMCEE \citep{Goodman10, Foreman-Mackey13} to characterize posterior probability. For each star,  we run each chain using 100 walkers and 100 steps, for a total of 10,000 random samples drawn from the posterior distribution. Because radial velocity doesn't depend on parallax and proper motion, we also sample 10,000 random samples for radial velocity and assume uniform priors on radial velocity. Therefore, for each star, we obtained 10,000 posterior samples including its heliocentric distance ($d$), tangential speed ($v$), direction of travel ($\phi$), and radial velocity ($rv$).  These random samples are used to derive Cartesian Galactocentric coordinate ($x,y,z$), the projected distance from the Galactic center, Galactic velocity components $(U, V, W)$, and Galactocentric cylindrical component $V_{\varphi}$ as described in section 2.2.   The median is used as an estimator of these astrophysical quantities, and the standard deviation of the quantities is used to define the uncertainty in the estimated value.
\par 
\cite{Bailer-Jones15} discussed in detail distance derivation by inverting the parallax is not appropriate when the relative parallax error is above 20 percent.
So we also compare distances of our sample stars with Sanders18 catalog \citep{Sanders18}. We cross-match between LAMOST DR5, Gaia DR2, and Bailer-Jones18 catalog. \cite{Bailer-Jones18} inferred distances for all 1.33 billion stars with parallaxes in Gaia DR2 using a weak distance prior distribution that varies smoothly as a function of Galactic longitude and latitude according to a Galaxy model. Panel (a) of Figure \ref{figure_app0} shows that the distances of stars with a relative error in parallax $0<f \equiv \sigma_{\varpi}/(\varpi-\varpi_{\rm zp}) \leq 0.1$ can be determined precisely just by inverting the parallax. As shown in the panel (b) of Figure \ref{figure_app0}, for the stars with $f>0.1$, estimating distances by inverting the parallax may lead to deviation when the distance of the stars is above 2 kpc. Panel (c) of Figure \ref{figure_app0} compares distance estimates obtained by Bayesian inference with Sanders18 catalog. We notice that the Bayesian inference performs better than inverting the parallax for the stars with $f>0.1$. Therefore, in this work, we choose different methods to derive the distance for the stars with $0<\sigma_{\varpi}/(\varpi-\varpi_{\rm zp}) \leq 0.1$ and $\sigma_{\varpi}/(\varpi-\varpi_{\rm zp}) > 0.1$. 

\par
We also compare our distances and velocities with other works such as Bailer-Jones18 catalog \citep{Bailer-Jones18} and Sanders18 catalog \citep{Sanders18} in Figure \ref{figure_app1}.  Panel (a) of Figure \ref{figure_app1} compares the distance estimates  from \cite{Bailer-Jones18} (${\rm d_{Bailer}}$)  and  this work (${\rm d_{This\ work}}$).  From this panel, we see that the distances of our sample stars  are very well consistent with  the results in \cite{Bailer-Jones18}, $93\%$ of stars have  ${\rm |d_{This\ work}}-{\rm d_{Bailer}}|<0.1$ kpc,  $6.3\%$ of stars have  ${\rm 0.1<|d_{This\ work}}-{\rm d_{Bailer}}|<0.3$ kpc, and  $0.65\%$ of stars have ${\rm 0.3<|d_{This\ work}}-{\rm d_{Bailer}}|<0.5$ kpc (only few stars with ${\rm |d_{This\ work}}-{\rm d_{Bailer}}|>0.5$ ).  \cite{Sanders18} also presented a catalog of stellar  distances, masses, and ages for $\sim3$ million giant  stars from large spectroscopic surveys (including APOGEE DR14). We also cross-match between APOGEE DR14, Gaia DR2, and Sanders18 catalog (i.e. our initial high-resolution sample),  for a total of 155,355 stars. We  compare the radial distance, vertical height, and rotational velocity  of these stars with Sanders18 catalog in panel (b), (c), and (d) of Figure \ref{figure_app1}, and our distance and velocity estimates are basically consistent with the results in \cite{Sanders18}. These imply that our comparison does not show significant bias.

\section{Comparison of stellar parameters}
\par
\begin{figure*}
	\centering
	\includegraphics[width=1.0\textwidth]{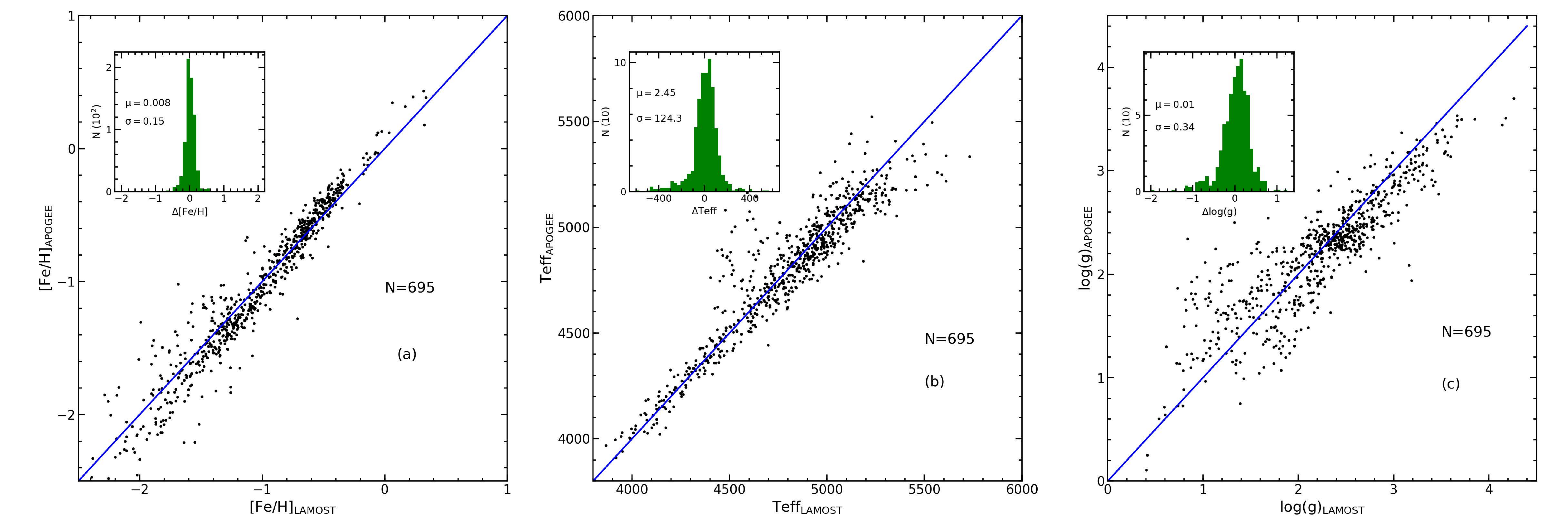}
	\caption{A comparison of metallicity ([Fe/H]), effective temperature (Teff), and surface gravity (log(g)) for the common stars between low-resolution and high-resolution samples. In the panel (a), ${\rm [Fe/H]_{LAMOST}}$ and ${\rm [Fe/H]_{APOGEE}}$ represent metallicity of the low-resolution and the high-resolution stars, respectively. $\Delta$[Fe/H] represent ${\rm [Fe/H]_{LAMOST}}$ $-$ ${\rm [Fe/H]_{APOGEE}}$. $\mu$ and $\sigma$ stand for mean and standard deviation of $\Delta$[Fe/H], respectively. The symbols in the panel (b) and (c) are also similar.}
	\label{figure_app2}
\end{figure*}

In this work, we have 695 common stars between high-resolution and low-resolution samples, and the comparisons of metallicity, effective temperature, and surface gravity for these common stars are given in Figure \ref{figure_app2}.   We also commented on the similarities and differences of the metallicity distribution functions based on the two samples. 
In order to validate LAMOST stellar parameters, there are several independent works comparing parameters of LAMOST with other reliable databases (including APOGEE) \citep[e.g.,][]{Wu11,Guo15,Xiang15,Luo15}. For example, \cite{Luo15} compared the parameters of the LAMOST with the APOGEE, and these parameters include  radial velocity, effective temperature, surface gravity, and metallicity. These comparisons do not show significant statistical deviation. \textbf{ For example, the mean difference of the metallicity between the LAMOST and the APOGEE is -0.09 dex, and the standard deviation is 0.08 dex.
The mean difference of the effective temperature between the LAMOST and the APOGEE is 1 K, and the standard deviation is 76 K. The mean difference of the surface gravity  is 0.08 dex, and the standard deviation is 0.28 dex. 
As shown in Figure \ref{figure_app2},  our comparison does not show significant bias, 
which is basically consistent with the results of \cite{Luo15}.   Although there are a small number of low-resolution stars that their parameters (e.g. surface gravity) deviate slightly from those of high-resolution stars in Figure \ref{figure_app2}, the influence of a single star could be ignored when the number of sample stars is large enough. } 

\par
\textbf{
It needs to be noted that the horizontal branch seems not distinct for the low-resolution sample in the bottom panel of Figure \ref{figure2}.  In order to check whether the accuracy of parameters effects on that, we also restrict the error of the effective temperature smaller than 70 K, the error of the surface gravity smaller than 0.1 dex, and signal-to-noise S/N $> 50$ in the $g$-band for the low-resolution sample. However,  the horizontal branch still seems not to be very distinct.  So we consider it could due to sample selection in the LAMOST survey. In addition, we only use the metallicity and radial velocity of the LAMOST sample in this study, the accuracy of temperature and surface gravity couldn't affect our final results. }

\end{document}